\documentclass[preprint2]{aastex631}

\def\Pab{Pa$\beta$}

\def\feii{Fe\,{\sc ii}}

\def\kms{km s$^{-1}$}

\def\1h07{1H\,0707$-$495}
\usepackage{ulem}

\def\aap{A\&A\/}
\def\apjs{ApJS}

\def\apjl{ApJL}

\begin{document}

\title{First detection of outflowing gas in the outskirts of the broad line region in 1H-0707-495\footnote{Based on observations obtained at the Southern Astrophysical Research (SOAR) telescope, which is a joint project of the Ministério da Ciência, Tecnologia e Inovações do Brasil (MCTI/LNA), the US National Science Foundation’s NOIRLab, the University of North Carolina at Chapel Hill (UNC), and Michigan State University (MSU).}}

\author[0000-0002-7608-6109]{Alberto Rodr\'{\i}guez-Ardila}
\affiliation{Laborat\'orio Nacional de Astrof\'isica (LNA)\\ Rua dos Estados Unidos 154, Bairro das Na\c c\~oes\\ CEP 37504-364, Itajub\'a, MG, Brazil}
\affiliation{Divis\~ao de Astrof\'isica, Instituto Nacional de Pesquisas Espaciais (INPE)\\ Avenida dos Astronautas 1758, S\~ao Jos\'e dos Campos\\ 12227-010, SP, Brazil}

\author[0000-0002-7865-3971]{Marcos Antonio Fonseca-Faria }
\affiliation{Laborat\'orio Nacional de Astrof\'isica (LNA)\\ Rua dos Estados Unidos 154, Bairro das Na\c c\~oes\\ CEP 37504-364, Itajub\'a, MG, Brazil}

\author[0000-0003-4153-4829]{Denimara Dias dos Santos}
\affiliation{Divis\~ao de Astrof\'isica, Instituto Nacional de Pesquisas Espaciais (INPE)\\ Avenida dos Astronautas 1758, S\~ao Jos\'e dos Campos\\ 12227-010, SP, Brazil}
\affiliation{Istituto Nazionale di Astrofisica (INAF)\\ Osservatorio Astronomico di Padova, 35122 Padova, Italy}

\author[0000-0002-5854-7426]{Swayamtrupta Panda}
\affiliation{Laborat\'orio Nacional de Astrof\'isica (LNA)\\ Rua dos Estados Unidos 154, Bairro das Na\c c\~oes\\ CEP 37504-364, Itajub\'a, MG, Brazil}

\author[0000-0001-9719-4523]{Murilo Marinello}
\affiliation{Laborat\'orio Nacional de Astrof\'isica (LNA)\\ Rua dos Estados Unidos 154, Bairro das Na\c c\~oes\\ CEP 37504-364, Itajub\'a, MG, Brazil}



\begin{abstract}

We use near-infrared (NIR) spectroscopy covering simultaneously the $zJHK$ bands to look for outflowing gas from the nuclear environment of \1h07\ taking advantage that this region is dominated by low-ionization broad line region (BLR) lines, most of them isolated. We detect broad components in H\,{\sc i}, Fe\,{\sc ii} and O\,{\sc i}, at rest to the systemic velocity, displaying full width at half maximum (FWHM) values of $\sim$500~\kms, consistent with its classification as a narrow-line Seyfert~1 AGN. Moreover, most lines display a conspicuous blue-asymmetric profile, modeled using a blueshifted component, whose velocity shift reaches up to $\sim$826~\kms. This last feature can be interpreted in terms of outflowing gas already observed in X-ray and UV lines in \1h07\ but not detected before in the low-ionization lines. We discuss the relevance of our findings within the framework of the wind scenario already proposed for this source and suggest that the wind extends well into the narrow line region due to the observation of a blueshifted component in the forbidden line of [S\,{\sc iii}]~$\lambda$9531.

\end{abstract}



\section{Introduction} \label{sec:intro}

Narrow-line Seyfert 1 galaxies (NLS1) are identified by their optical emission line properties \citep{osterbrock+85, goodrich89, Leighly_1999b, Mathur_2000, Sulentic_etal_2000, Veron-Cetty_etal_2001, Collin_Kawaguchi_2004, Zhou_etal_2006, Komossa_Xu_2007, Woo_etal_2015, Rakshit_etal_2017}. Among their main characteristics, we highlight the presence of narrow permitted optical lines (full width at half maximum of the broad component of H$\beta$ $<$ 2000~km\,s$^{-1}$) and weak forbidden lines ([O\,{\sc iii}]/H$\beta <$ 3; this distinguishes them from Seyfert 2 galaxies). Moreover, they frequently show strong Fe{\sc ii} emission (R$_{4570} >$ 1\footnote{The strength of \feii{}, R$_{\rm4570}$, is gauged by computing the ratio of the integrated \feii{} intensity/flux between 4434\,$-$\,4684~\AA, denoted as \feii{} $\lambda$\,4570, to the flux of the broad component of the H$\beta$\
%footnote{The \feii{} intensity is determined through the EW ratio as EW(\feii{} $\lambda$\,4570\,$\AA$)/EW\,(H$\beta$)} \swayam{the second footnote doesn't appear in the compiled PDF. Also, this point seems redundant since we say in Footnote 1 that R$_{4570}$ is the flux/intensity ratio.} 
such that R$_{\rm4570}$\,\,$\approx$\,F(\feii{} $\lambda$\,4570\,$\AA$)/F\,(H$\beta$).}) 
\citep{osterbrock+85,goodrich89}. Another characteristic that makes these sources interesting is their black hole mass, which is systematically smaller (M$_{\rm BH} < 10^8$~M$\odot$) than that in broad line AGN. This implies a higher accretion rate relative to the Eddington value compared with Seyfert~1 galaxies with broad optical lines \citep[See][for a comprehensive review]{komossa18}. A special subset of this class, known as extreme NLS1s or xA sources within the Main Sequence classification \citep{marziani+18,panda+19,marinello+20} shows highly super-Eddington accretion sources \citep{jin+17,panda+23}. These xA objects exhibit spectral features such as extremely prominent Fe\,{\sc ii} (i.e., R$_{4570} > $1.5) \citep{wang14, du2015, marinello+20}, very rapid X-rays and UV flux variability \citep{gallo06}. In some cases, they also harbor Ultrafast outflows (UFOs) signatures \citep{tombesi+10,parker+17}.

The current understanding of UFOs is that they are powerful winds from the AGN with outflow velocities larger than 10000~km\,s$^{-1}$. They are usually identified in the X-rays utilizing high-energy absorption features from FeXXV/XXVI in the 7–10 keV energy band \citep[e.g.,][]{tombesi+10}.  It has been suggested that UFOs originate in winds magnetically or radiatively driven off the AGN accretion disc at high Eddington rates \citep{pounds+03,reeves+03,fukumura+15}. If this scenario is correct, these winds are of great interest because they are very strong candidates for driving AGN feedback, as they couple with galactic gas much more efficiently than jets.

\1h07\ is a low-redshift (z=0.04) NLS1 galaxy, well known for its extreme variability and
spectral shape \citep[e.g.,][]{turner+99,leighly99,boller+02}. Moreover, it displays a very strong soft X-ray excess and relativistic broad iron line, thought to arise from emission reprocessed by the accretion disk \citep{fabian+09}. \1h07\ was also the first AGN where an X-ray reverberation lag was detected \citep{fabian+09,zoghbi+10}. One characteristic that makes this AGN interesting is the presence of blueshifted absorption features from a UFO, detected by \citet{dauser+12} and \citet{hagino+16} using XMM-Newton spectra. Moreover, in its X-ray spectra emission lines from O\,{\sc viii} and N\,{\sc vii} are observed \citep{kosec+18}, in particular, when the continuum flux is low. Various authors have at times invoked low ionization partial-covering absorption in \1h07, either as a way of producing the spectral structure at 7~keV at the Fe~K edge \citep[e.g.,][]{mizumoto+14} or as a way of producing soft X-ray variability \citep[e.g.,][]{boller+21}.

In the ultraviolet (UV), Hubble Space Telescope (HST) STIS (Space Telescope Imaging Spectrograph) observations of \1h07\ reported by \citet{leighly+04} shows that at least part of the outflows detected in the X-rays is also observed in the UV emission lines. Overall, the STIS spectrum is characterized by a very blue continuum; broad, strongly blueshifted high-ionization lines (including C\,{\sc iv} and N\,{\sc {v}}); narrow, symmetric intermediate- (including C\,{\sc iii}], Si\,{\sc iii}], and Al\,{\sc iii}) and low-ionization (e.g., Mg\,{\sc ii}) lines. These latter features are centered at the rest wavelength. The study of their emission-line profiles reveals that the high-ionization lines are associated with a wind while the intermediate- and low-ionization lines arise in low-velocity gas. This latter component is likely associated with the accretion disk or with the base of the wind.

Despite the wealth of information about \1h07\ in the X-ray, UV, and optical, very little is known about the near-infrared (NIR) properties of this source. Indeed, to the best of our knowledge, only \citet{durre+22} employed this AGN as part of a southern sample of NLS1 to study the kinematics of the broad line region (BLR) using the Pa$\alpha$ line. The NIR wavelength range, though, is highly valuable because (i) the \feii{} emission lines, notably the \feii{} lines around the 1-$\mu$m region, are isolated or semi-isolated, unlike in the UV-optical region, allowing a better and more accurate determination of the line properties \citep{rudy+00,rodriguez+02,riffel+06}; (ii) the H\,{\sc i} lines, particularly, Pa$\alpha$ and Pa$\beta$ are isolated, allowing the characterization of their emission line profiles; (iii) continuum emission due to hot dust starts to show up in this spectral region.

In this work, we report near-infrared (NIR) spectroscopy carried out on 1\1h07\ aimed at studying the outflow properties already detected in the high-ionization BLR gas but up to today elusive to detection in low-ionization BLR lines. 
This work is organized as follows: Section~\ref{obs_data} provides an overview of the data and observations. Section~\ref{results} presents the results and in Section~\ref{discussion} a discussion and interpretation of the findings are made. Finally, in Section~\ref{conclusion} we give the main conclusion found from the data analysis.
%%%%%%%%%%%%%%%%%%%%%%%%%%%%%%%%%%%%%%%%%%
\section{Observations and data reduction}\label{obs_data}

NIR spectroscopy of \1h07\ was obtained using the Triplespec4 spectrograph \citep{schlawin+14} attached to the 4.1\,m Southern Astrophysical Research Telescope (SOAR) on the night of 23 February 2022. The science detector employed is a 2048 $\times$ 2048 Hawaii-2RG Hg-Cd-Te array with a sampling of 0.41 arcsec/pixel. The slit assembly is 1.1 arcsec wide and 28 arcsec long. The delivered spectral resolution R is $\sim$2850 across the different dispersion orders. Observations were done nodding in two positions along the slit. Right before the science target, the A1V star HIP\,32913 (V=7.57), close in airmass to the former (airmass = 1.06), was observed to remove telluric features and to perform the flux calibration. Wavelength calibration was carried out using skylines present across the NIR spectra.

The spectral reduction, extraction, and wavelength calibration procedures were performed using {\sc spextool v4.1}, an IDL-based software developed and provided by the SpeX team \citep{cushing+04}  with some modifications specifically designed for the data format and characteristics of ARCoIRIS, written by Dr. Katelyn Allers (private communication). Telluric feature removal and flux calibration were done using {\sc xtellcor} \citep{vacca+03}. The different orders, extracted using an aperture window of 2" centered at the peak of the source light profile, were merged into a single 1D spectrum from 0.94 to 2.4~$\mu$m using the {\sc xmergeorders} routine.  The final reduced spectrum includes an error vector and appears as a third extension in the data file. It measures the uncertainty in flux calibration at every wavelength and takes into account errors propagated through the extraction process. That extension is employed to estimate the errors associated with the determination of the integrated flux of the lines and continuum fitting.

The final, merged spectrum was corrected for redshift, determined from the brightest lines detected. We employed Pa$\alpha$, Pa$\beta$, O\,{\sc i}~1.128~$\mu$m, Fe\,{\sc ii}~1.0501~$\mu$m and [S\,{\sc iii}]~0.953~$\mu$m to this purpose. The average value obtained was $z=0.04126$, in excellent agreement with the values reported in the literature \citep{leighly+97,leighly+04}  and in the NASA/IPAC Extragalactic Database. Afterward, we corrected the spectrum for Galactic extinction using the Cardelli et al. law \citep{cardelli+89} and the extinction maps of \citet{schlafly+11}. A value of E(B-V)= 0.084 was adopted.

\begin{figure*}[]
\centering
\includegraphics[width=0.9\textwidth]{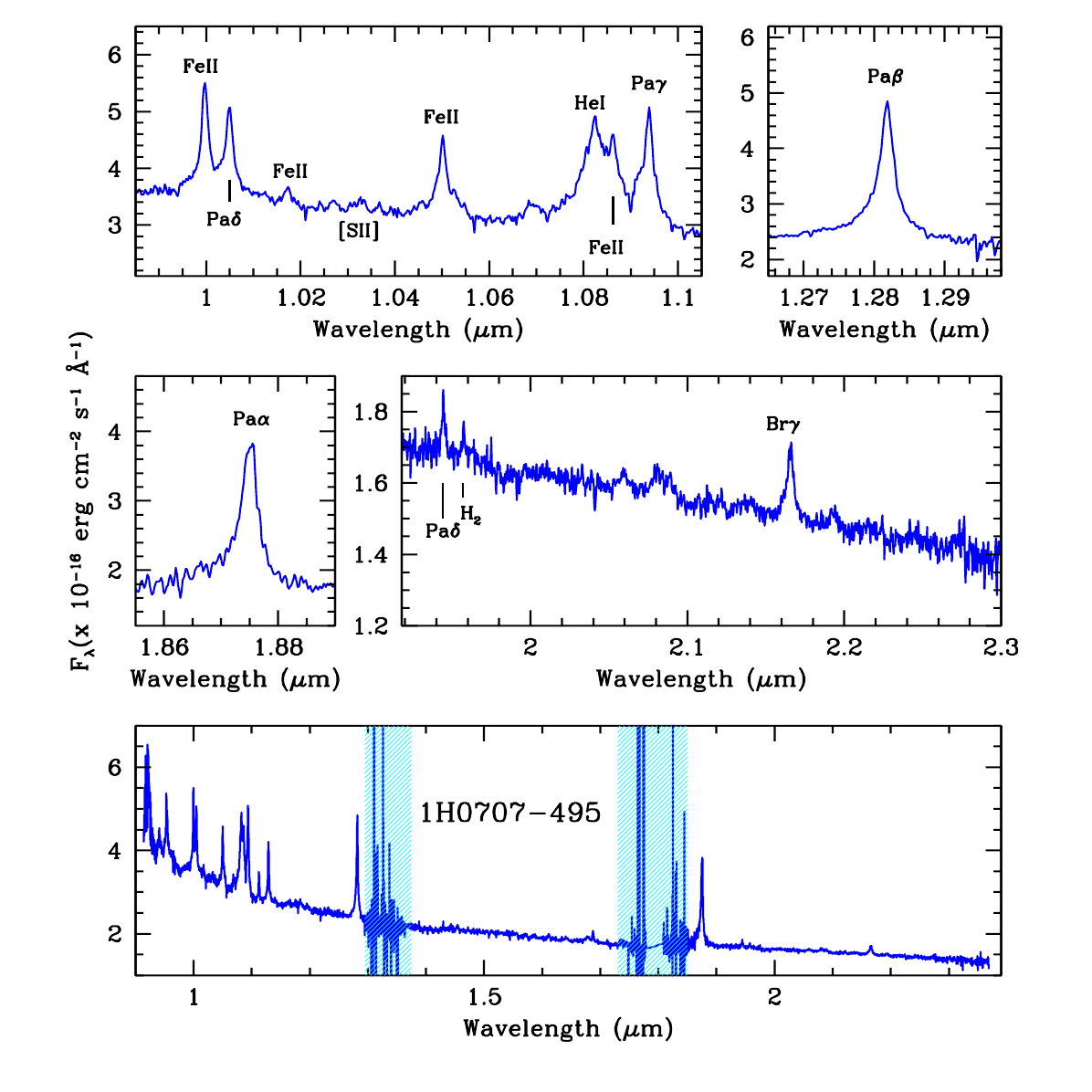}
\caption{Near-IR spectrum of 1H\,0707-495 in rest wavelength already corrected for a Galactic extinction E(B-V)=0.084. The bottom panel shows the spectrum in the wavelength interval 0.94$-$2.4~$\mu$m while the boxes in the upper and mid-rows are insets of the bottom spectrum at some spectral regions of interest to this work. The most important emission lines are identified. The shaded areas in blue are regions of bad atmospheric transmission. \label{fig:NIRspec}}
\end{figure*}

Figure~\ref{fig:NIRspec} displays the most relevant characteristics observed in the redshifted and Galactic extinction corrected NIR spectrum of \1h07. To the best of our knowledge, this is the first time in the literature that a NIR spectrum of this AGN covering simultaneously the 0.94 $-$ 2.4~$\mu$m interval in rest-wavelength is presented. The middle and upper insets display the region around the 1$\mu$m~Fe\,{\sc ii} lines, Pa$\beta$, Pa$\alpha$, and the $K-$band region. The most important emission lines are identified.  

\unskip

%%%%%%%%%%%%%%%%%%%%%%%%%%%%%%%%%%%%%%%%%%
\section{Results}\label{results}

Overall, the NIR spectrum of 1H\,0707-495 displays similar characteristics to that observed in other NLS1 \citep{rodriguez+02,riffel+06,marinello+16,marinello+20}. The most important difference is the lack of bright emission features from the narrow-line region (NLR), such as the forbidden [S\,{\sc iii}]~0.953~$\mu$m, [Fe\,{\sc ii}]~1.257~$\mu$m, molecular lines from H$_2$, and coronal lines, which are rather faint or not detected in \1h07. This result is consistent with previous observations \citep{paul+21}. Moreover, the NLR component of the permitted H\,{\sc i} lines seems absent, except likely in the Paschen lines. 

The continuum emission is dominated by a strong featureless component, with the flux increasing steeply towards shorter wavelengths. No evidence of stellar absorption features were detected in the observed spectrum. Thus, we assume that the stellar contribution is below 10\% of the observed continuum \citep{riffel+06}.

\begin{figure*}[]
\centering
\includegraphics[width=0.9\textwidth]{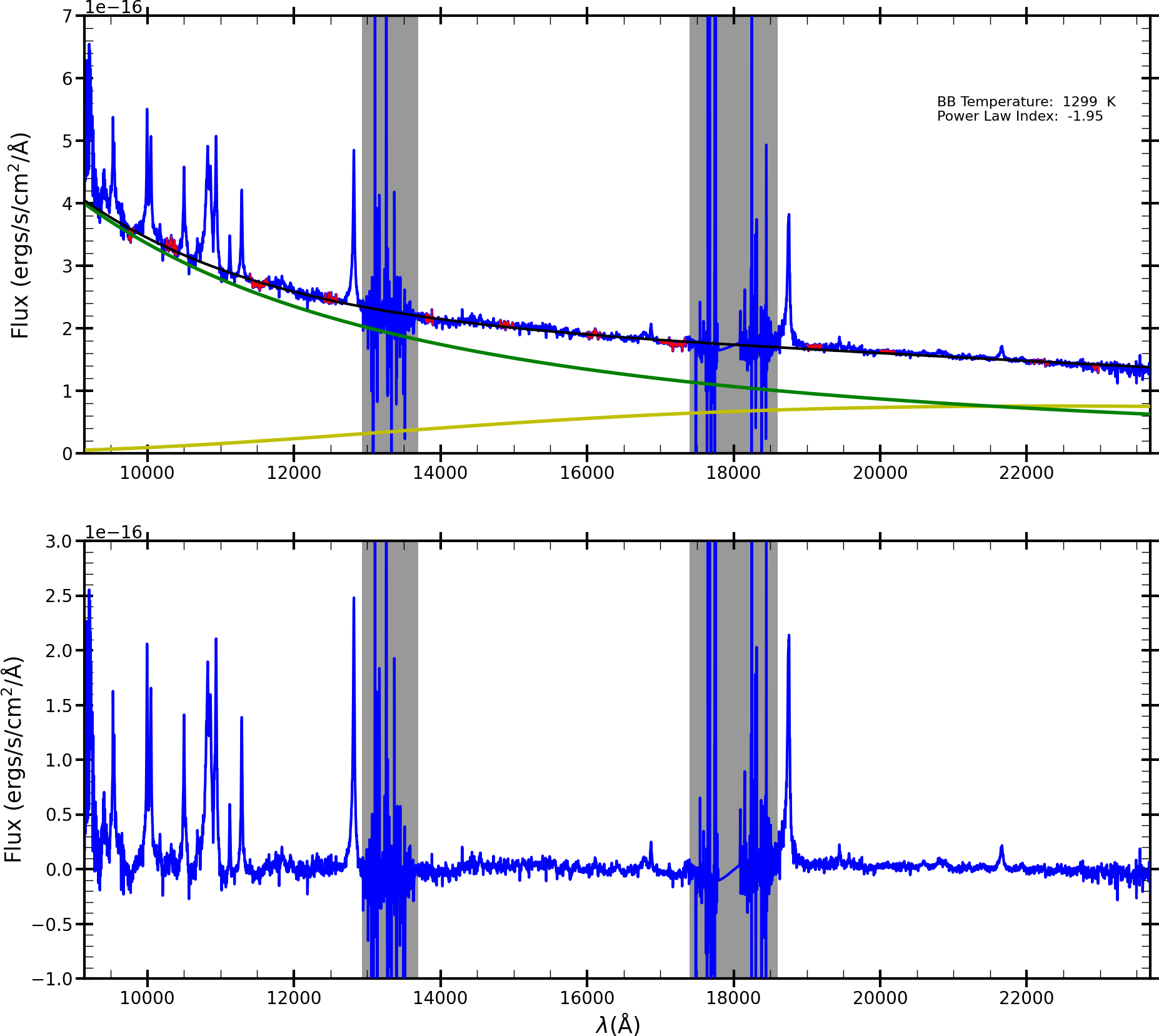}
\caption{Upper panel: Fit applied to the NIR continuum emission (black line) in 1H\,0707-495 with a composite function, consisting of a power-law (green line) of the form F$_{\lambda} \propto \lambda^{\alpha}$ with $\alpha$=-1.95 and a blackbody (gold line) of temperature $T_{\rm BB}$=1300~K. The spectral windows employed in the fitting are marked in red. The bottom panel shows the nebular spectrum after subtraction of the continuum emission. The grey stripes are regions of bad atmospheric transmission. \label{fig:cont_fit}}
\end{figure*}

To characterize the NIR continuum emission, we first fit a function described by a power-law of the form F$_{\lambda} \propto \lambda^{\alpha}$, where $\lambda$ is the wavelength and $\alpha$ the spectral index of the power-law. Special care was taken to exclude emission lines in the spectral windows that were used in the fit. We found unsatisfactory results with that function and concluded that it alone cannot reproduce the observed NIR continuum. We then employed a composite function, consisting of an underlying power-law plus a blackbody component. It has been employed successfully in the fitting of the NIR continuum in other AGNs, including NLS1 \citep{glikman+06,riffel+09,landt+11}. Our results show that a power-law with spectral index $\alpha=-1.95\pm0.25$ and a blackbody of temperature $T_{\rm BB}=1299\pm100$~K reproduce satisfactorily the observed continuum. The errors quoted for the aforementioned parameters are within 2$\sigma$ uncertainties. Figure~\ref{fig:cont_fit} shows the fit (upper panel) and the nebular spectrum after subtraction of that component. 

The value of $\alpha$  derived in this work in the NIR is, within errors, very close to that of -2.3 reported by \citet{leighly+04} for the UV/optical continuum. We note that these latter authors do not quote any uncertainty associated with the spectral index. Despite this, we conclude that the power law in the NIR likely represents the low-energy tail of the continuum produced by the central source. The blackbody component is associated with the hottest dust component, likely located in the inner face of the obscuring torus \citep{glikman+06,landt+11}. 

After the continuum fitting, we subtracted the power-law and the blackbody components from the observed spectrum. This procedure is carried out to study the pure emission line spectrum of \1h07. However, the analysis that will be presented in Section~\ref{sec:spectrum1h07} is not affected by this step (see below).

\unskip

\subsection{The emission line spectrum of 1H\,0707-495}
\label{sec:spectrum1h07}

The emission line spectrum in 1H\,0707-495 is dominated by permitted, broad features, most of them associated with the BLR. The only NLR lines detected are the [S\,{\sc iii}]~0.953~$\mu$m in the blue edge and the H$_2$ line at 1.957~$\mu$m in the $K-$band. Both features are intrinsically faint. It is important to notice that the forbidden emission line of [S\,{\sc iii}] is usually the strongest forbidden line in the NIR \citep{riffel+06}. We characterized the observed emission lines in terms of the integrated flux, full-width at half maximum (FWHM), and centroid position of the line peaks. To this purpose, we fitted Gaussian or Lorentzian functions to individual lines or to sets of blended lines. This procedure was carried out using a set of custom 
scripts written in {\sc python} by our team. The code uses non-linear least squares, with {\sc scipy.optimize.curve\_fit}. Open source software was employed, such as {\sc MATPLOTLIB} \citep{hunter2007}, {\sc NUMPY} \citep{van2011}  and {\sc SCIPY} \citep{virtanen2020}. For each emission line, up to 3 Gaussian components or 2 Gaussian components and one Lorentzian are allowed. Constraints for the lines were also added such that lines belonging to the same ion should have the same FWHM (in velocity space) and obey the theoretical wavelength separation.

\begin{figure*}
\includegraphics[width=9cm]{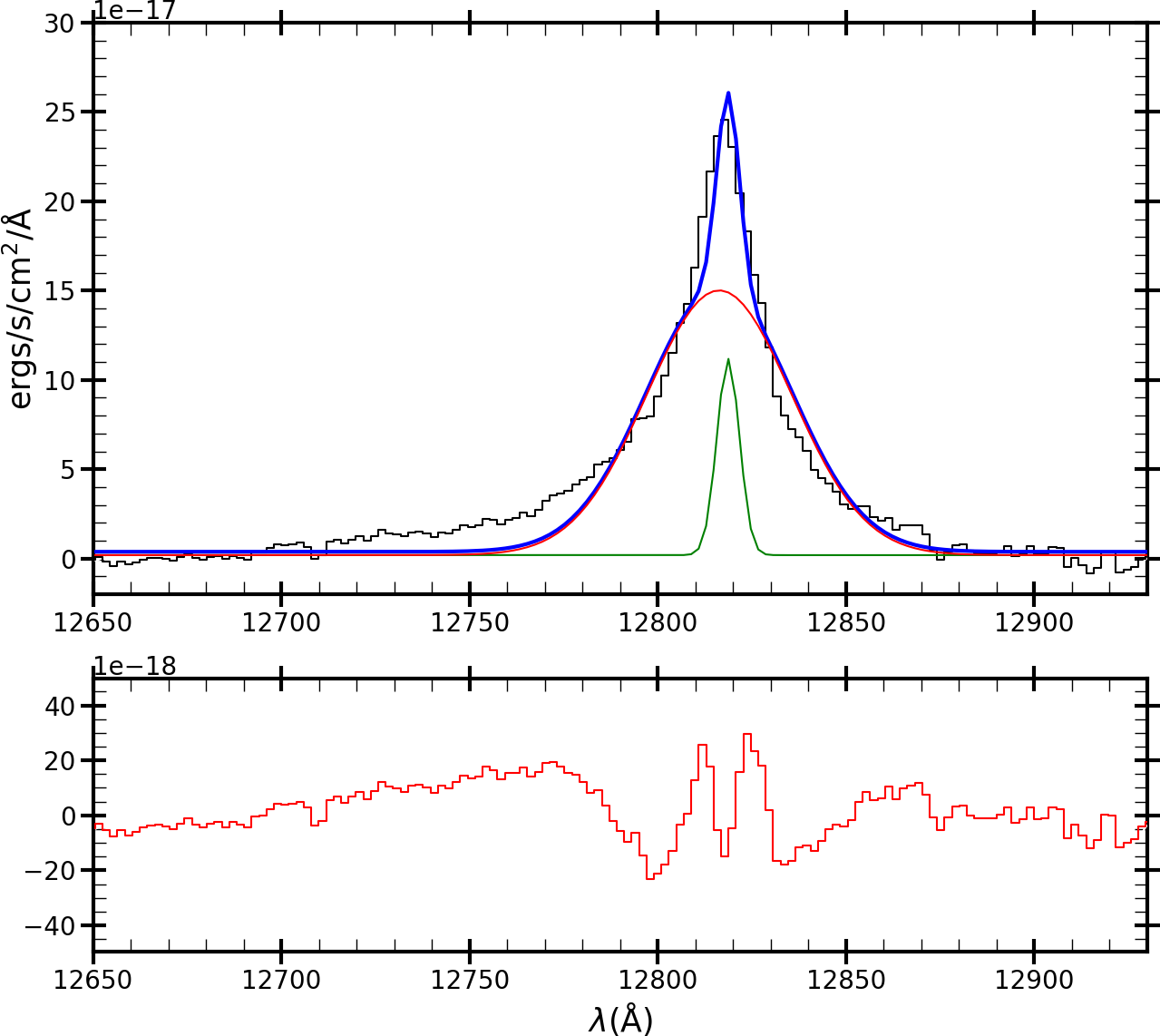}
\includegraphics[width=9cm]{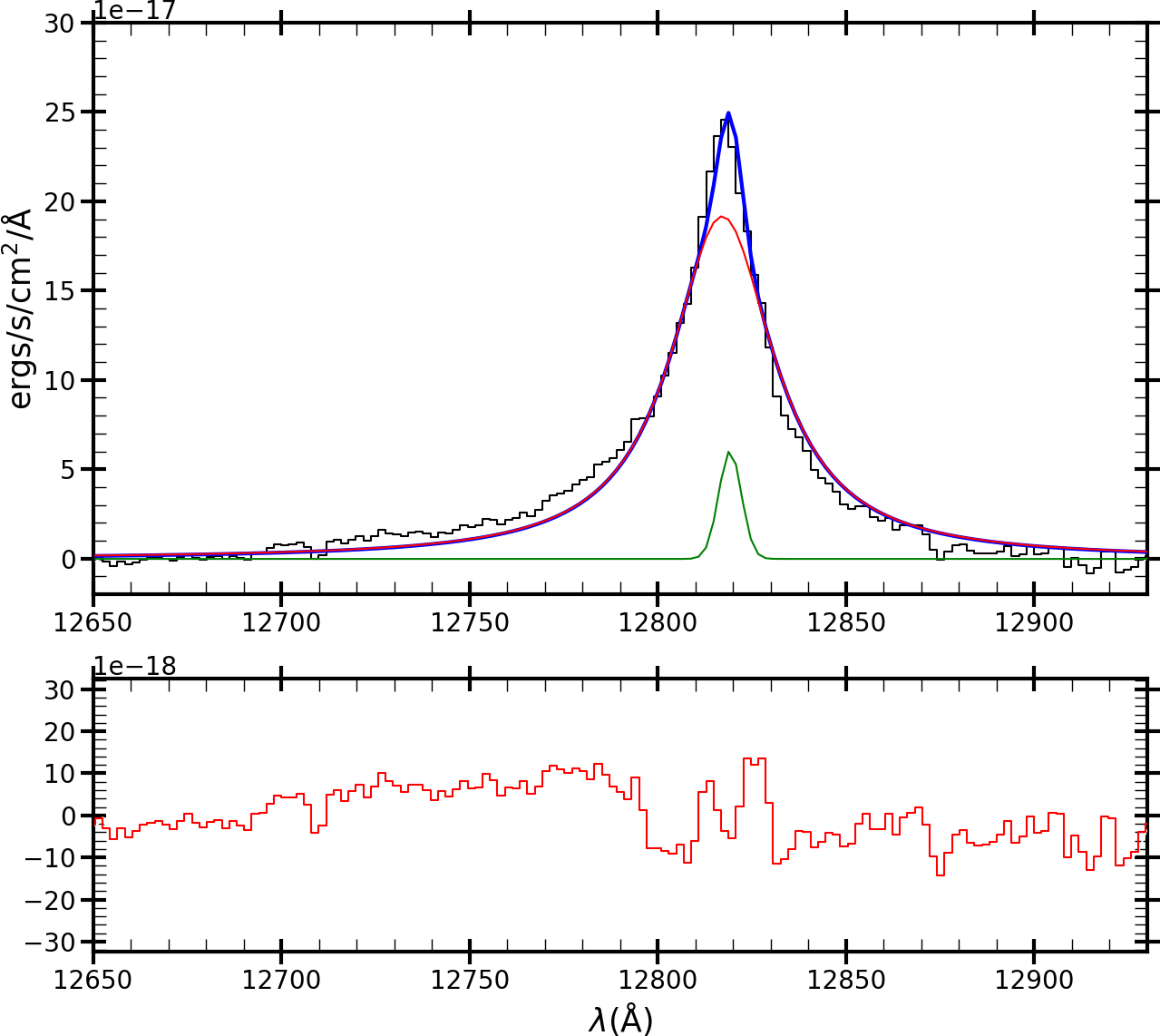}
\includegraphics[width=9cm]{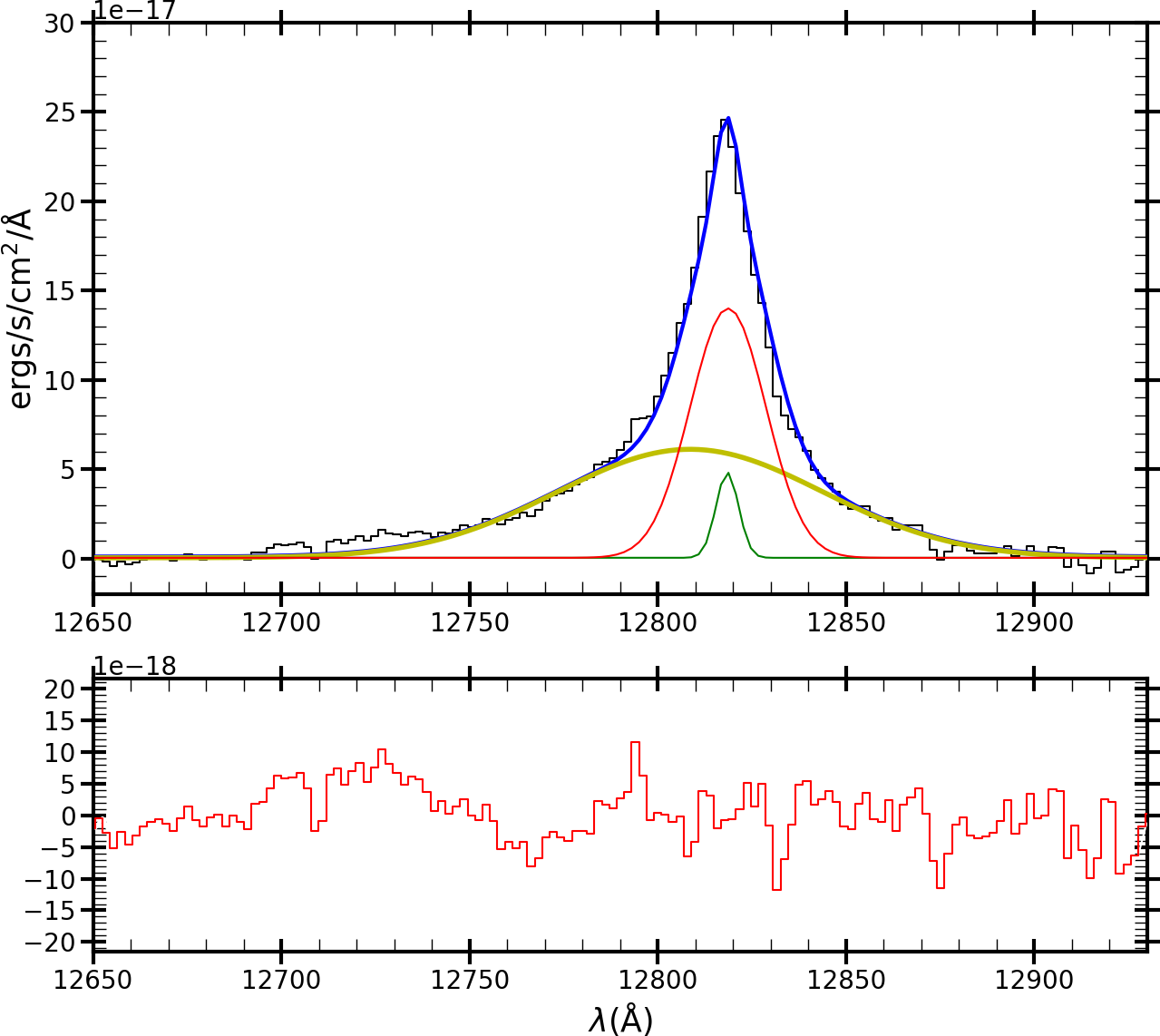}
\includegraphics[width=9cm]{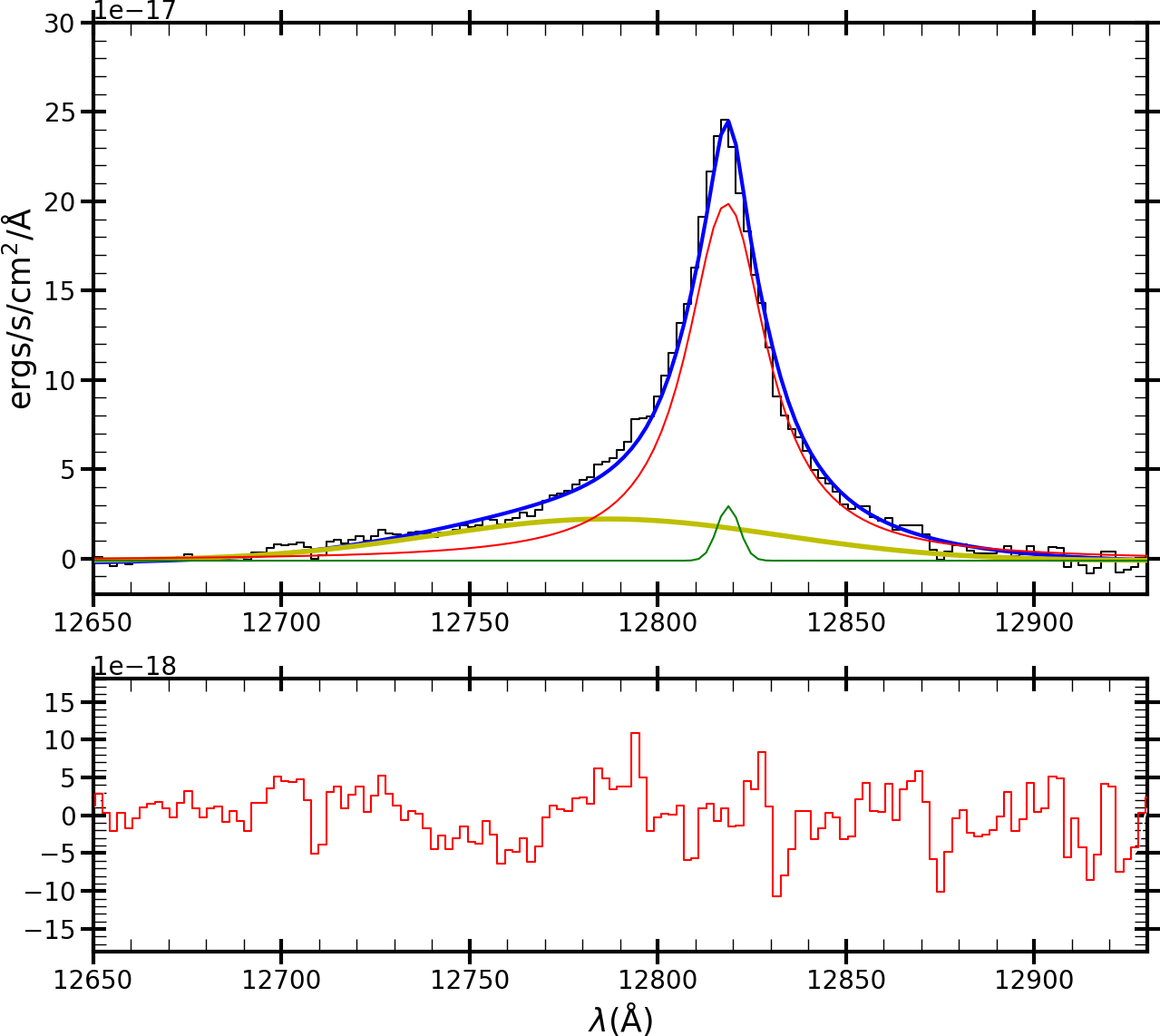}
\caption{Line fit carried out to the Pa$\beta$ line. The upper two panels show the results after considering a BLR composed of a single Gaussian (left) or a Lorentzian component (right). The BIC found from this approach are in Table~\ref{tab:BIC_results}. The bottom two panels display the results after the addition of a blueshifted Gaussian component to the classical Gaussian BLR (left) and a Lorentzian BLR (right). The corresponding BIC values are also in Table~\ref{tab:BIC_results}. 
In all panels, the observed profile is the black histogram, the green curve is the NLR contribution, the red curve is the BLR contribution and the yellow curve is the additional blueshifted Gaussian component associated with the BLR. The blue line is the modeled profile. The bottom panel is the residual after subtracting the modeled emission line profile from the observations.   \label{fig:Pabeta}}
\end{figure*}

Figure~\ref{fig:Pabeta} shows a zoom around the region where the Pa$\beta$ line is located. It is the strongest BLR line detected in the NIR spectrum. 
We first fit the observed profile with two components - a narrow one, from the NLR and a broad profile, associated with the BLR. The results are presented in the upper two panels of Figure~\ref{fig:Pabeta}. In the left panel, the BLR is represented by a single Gaussian profile while in the right panel, it is assumed that the BLR contribution can be represented by a Lorentzian profile. A quick inspection of both fits evidences residuals as large as $\sim$10\% of the peak intensity of the line. Moreover, the largest residuals are mostly concentrated towards the blue wing of the emission line profile.

We then tested a scenario of a BLR consisting of two components to account for the observed large residuals in the previous case. The results are shown in the bottom two panels of Figure~\ref{fig:Pabeta}. In both cases, we kept the former BLR component (Gaussian -- left panel, and Lorentzian, right panel) and included an additional blueshifted broad Gaussian component. All the parameters associated with this extra component were left free in the fit. It can be seen that after the addition of the latter component the residuals were considerably reduced in both cases, being limited now to $\sim$3\% of the line peak.

Although the addition of a second BLR component offers a better description of the observed Pa$\beta$ profile, it is well-known that the root-mean-square (RMS) of the continuum residuals decreases with the increase of the number of components, regardless of their physical meaning. However, it is worth noticing at this point that previous works on \1h07\ had already reported the detection of blue-asymmetric profiles in UV lines such as C\,{\sc iv}, Si\,{\sc iv}, and O\,{\sc iv} \citep{leighly+04}. \citet{leighly+04} also point out that low-ionization lines such as Mg\,{\sc ii} in the UV or H$\beta$ in the optical were modeled using a Lorentzian profile, with no evidence for either blue or red asymmetries and centered at the systemic velocity. It is important to mention, though, that the latter two lines are located in regions that are strongly contaminated by broad humps of Fe\,{\sc ii} emission. This makes the full characterization of their line profiles very uncertain. 

Even though our results on Pa$\beta$ agree with previous observational findings using UV spectroscopy on this AGN, to have firm evidence of the need for a second component, we employed the Bayesian Information Criteria (BIC). It was first introduced by \citet{schwarz78} to guide model selection, which is the problem of distinguishing competing models, sometimes featuring different numbers of parameters. Here, we apply the BIC to test if a second Gaussian component is indeed necessary. According to \citet{schwarz78}, BIC is defined by Equation~\ref{eq:1},

\begin{equation}\label{eq:1}
    BIC = -2 * ln(L) + k * ln(N)
\end{equation}

where $L$ is the likelihood of the model, $k$ is the number of parameters, and $N$ is the number of observations.

Assuming Gaussian errors and the boundary condition that the derivative of the log likelihood with respect to the true variance is zero, according to \citet{liddle07} Equation~\ref{eq:1} can be expressed in terms of the residual sum of squares (RSS):

\begin{equation}\label{eq:2}
    BIC = N*ln(RSS/N) + k*ln(N)
\end{equation}

The model giving the smallest BIC among the candidates is the one favored. If the difference in BICs between two competing models is 0–2, this constitutes ``weak'' evidence in favor of the model with the smaller BIC; a difference in BICs between 2 and 6 constitutes ``positive" evidence; a difference in BICs between 6 and 10 constitutes ``strong" evidence; and a difference in BICs greater than 10 constitutes ``very strong" evidence in favour of the model with smaller BIC.

It is important to mention that the BIC attempts to mitigate the risk of over-fitting by introducing the penalty term $k * ln(N)$, which grows with the number of parameters. This allows us to filter out unnecessarily complicated models, which have too many parameters to be estimated accurately on a given data set of size N.

\begin{table*}[]
    %\centering
    \footnotesize
    \setlength{\tabcolsep}{4pt} % Default value: 6pt
    \begin{tabular}{lcccccc|cccccc}
    \hline \hline
     & \multicolumn{6}{c}{Gaussian BLR} &   \multicolumn{6}{c}{Lorentzian BLR} \\
    \cline{2-13}  
           &    &   &     &    &     &   &   &     &  &   &  &      \\  
           Line &  N &  K$^{*}$ & K$^{**}$ &   BIC$^{*}$ & BIC$^{**}$ & $\Delta$\,BIC  &  N &  K$^{*}$ & K$^{**}$ &   BIC$^{*}$ & BIC$^{**}$ & $\Delta$\,BIC  \\ 
           &    &   &     &    &     &   &   &     &  &   &  &      \\  
     %&  &   &    &  &  ( BIC$^{**} -$ BIC$^{*}$  ) &  &   &  & & ( BIC$^{**} -$ BIC$^{*}$  )  \\ 

    \hline
     % LINHAS 1 %%%%%%%%%%%%%%%%%%%%%%%%%%%%%%%%%%%%%%%%%%%55
     Pa$\beta$    
      &  87 & 6 & 9 & -6747 &  -6910 &  163  &
         87 & 6 & 9 &  -6836 & -6938  &  102  \\
           &    &   &     &    &     &   &   &     &  &   &  &      \\  
    \hline
     % LINHAS 2 %%%%%%%%%%%%%%%%%%%%%%%%%%%%%%%%%%%%%%%%%%%55
     \feii    
      &  68 & 12 & 15 &  -5288 &  -5359 &  71  &
         68 & 12 & 15  &  -5339 &  -5377 &    38   \\
      $\lambda$10501     &    &   &     &    &     &   &   &     &  &   &  &      \\  
         &    &   &     &    &     &   &   &     &  &   &  &      \\  

    \hline
     % LINHAS 3 %%%%%%%%%%%%%%%%%%%%%%%%%%%%%%%%%%%%%%%%%%%55
     O\,{\sc i}    
      &  72 & 3 & 6  &  -5649  &  -5733 &  84  &
         72 & 3 & 6 &  -5769 & -5721   &   -48   \\
           &    &   &     &    &     &   &   &     &  &   &  &      \\  
      
     \hline
     % LINHAS 4 %%%%%%%%%%%%%%%%%%%%%%%%%%%%%%%%%%%%%%%%%%%55
     \feii+    
      &  138 & 15 & 21  &  -10690 &  -10938 &  248  &
         138 & 15  & 21  & -10848 & -10931 &  83   \\
      Pa$\delta$+     &    &   &     &    &     &   &   &     &  &   &  &      \\  
      He\,{\sc ii}+   &    &   &     &    &     &   &   &     &  &  &  &       \\  
       \feii     &    &   &     &    &     &   &   &     &  &    &  &     \\  

     \hline
     % LINHAS 5 %%%%%%%%%%%%%%%%%%%%%%%%%%%%%%%%%%%%%%%%%%%55
    {[}S\,{\sc iii}{]}+     
      &  53 & 15& 21 &    -4104 & -4170  &  66  &
         53 & 15 & 21   &  -4090 &  -4161 &    71   \\  
      Pa\,8     &    &   &     &    &     &   &   &     &  &   &  &      \\       
    \hline \\
    \end{tabular}
    \caption{Parameters employed in the line fitting procedure and resulting BIC values. N is the number of data points, K is the number of parameters modeled (three for every Gaussian or Lorentzian function) and BIC is the Bayesian information criterion determined by Equation~\ref{eq:1}. Columns 2-7 list the results when a classical Gaussian BLR is assumed and columns 8-13 show the results for a Lorentzian BLR. 
    $^{*}$ values without a second Gaussian component; $^{**}$ values when considering a second, blueshifted Gaussian component. $\Delta$\,BIC is the difference of BIC for model $i$ and the minimum BIC value, BIC$_{\rm min}$. }
    \label{tab:BIC_results}
\end{table*}

Using Equation~\ref{eq:2} and the values of $N$ and $k$ listed in Table~\ref{tab:BIC_results},  we determine the BIC for the profile modeling applied to Pa$\beta$ and depicted in Figure~\ref{fig:Pabeta}. The results are in Columns 5-6 and 11-12 of that same Table. It is possible to see that the smallest BICs (-6910 and -6938) are obtained when the BLR is modeled with two components: a broad profile (Gaussian or Lorentzian) with the line peak very close to the systemic velocity (the classical BLR) plus a blueshifted, broad Gaussian component. Moreover, the difference  $\Delta$BIC  (Columns 7 and 13) between the BLR represented by one or two components is $>$10 regardless of the profile employed to represent the classical BLR. 
Therefore, we conclude that a broad, blueshifted, component is a requirement to properly model the observed Pa$\beta$ profile. Here, we suggest that this extra component represents the outflow already detected in UV lines by \citet{leighly+04} in this object. 

Table~\ref{tab:line_fit} lists the centroid position, flux, full-width at half maximum (FWHM), and the shift of the line peak relative to the laboratory position found for each component of both the Gaussian and the Lorentzian approach. The errors in the fluxes are representative at the 2$\sigma$ level across all cases. 
 
The BIC values from the Pa$\beta$ fit also allow us to test whether the classical BLR is best described by a Gaussian or a Lorentzian profile. We found that the smallest BIC is obtained with the latter function. The $\Delta$BIC between these two models (Columns 7 and 13) is 28, that is, larger than 10. This agrees to previous results reported by \citep{leighly+04} to model the H$\beta$ line.

When a Lorentzian profile is employed to represent the classical BLR,  it has an FWHM of 602~\kms\ and is centered at the systemic velocity. The outflow component, modeled using a Gaussian component, is found to be blueshifted by -725~\kms\ and has a FWHM of 2545~\kms. We note that the maximum error associated with the peak position of the line is 30~\kms\ while that associated with the FWHM is 35~\kms. These values were extracted from the maximum value of the root mean square (RMS) of the wavelength calibration provided by the reduction pipeline, and from the measurement of the FWHM around the Pa$\beta$ line, assuming different continuum levels, respectively. They can be considered as standard for the analysis presented in the remainder of this work.

We notice that the Pa$\alpha$ line in the $K-$band is also detected in our data (see Figure~\ref{fig:NIRspec}). However, we refrain from using it in the analysis done here because the blue wing of the line is very close to the poor transmission region between the $H-$ and $K-$ bands. An inspection of Figure~\ref{fig:NIRspec} allows us to confirm the presence of a strong blue-asymmetric profile in Pa$\alpha$, very similar to that of \Pab. However, because the former displays a lower S/N  and is fainter than the latter (which contradicts the Paschen decrement), we opted to leave Pa$\alpha$ out.

\begin{table*}[]
    %\centering
    \footnotesize
    \setlength{\tabcolsep}{2.5pt} % Default value: 6pt
    \renewcommand{\arraystretch}{1.5} % Default value: 1
    \begin{tabular}{|l|c|c|c|c|cc|c|c|c|}
    \hline
     & \multicolumn{4}{|c|}{Gaussian BLR} & & \multicolumn{4}{c|}{Lorentzian BLR} \\
     \cline{2-5} \cline{6-10}
    Line & Centre & Flux & FWHM & $\Delta\lambda$ & & Centre & Flux & FWHM & $\Delta\lambda$ \\
     & (\AA) &  & (\kms) & (\kms) & & (\AA) & & (\kms) & (\kms) \\
     \hline
     {[}S\,{\sc iii}{]}$_{\rm ~NLR}$ & 9532 & 0.49$\pm$0.04 & 122 & 41 & & 9532 & 0.47$\pm$0.04 & 119 & 41 \\
      {[}S\,{\sc iii}{]}$_{\rm ~out}$ & 9526 & 0.83$\pm$0.13 & 565 & -170 & & 9526 & 0.78$\pm$0.14 & 583 & -164 \\
       Pa\,8$_{\rm ~NLR}$ & 9546 & 0.10$\pm$0.05 & 118 & 0 & & 9546 & 0.08$\pm$0.04 & 118 & 0 \\
       Pa\,8$_{\rm ~BLR}$ & 9547 & 1.18$\pm$0.16 & 550 & 28 & & 9547 & 2.33$\pm$0.15 & 597 & 16 \\
       Pa\,8$_{\rm ~out}$ & 9541 & 1.37$\pm$0.44 & 1934 & -157 & & 9525 & 0.83$\pm$0.64 & 2533 &  -669 \\
     \feii$_{\rm ~BLR}$ & 9997 &  2.11$\pm$0.13 & 408 & 9  &  & 9997 &  4.02$\pm$0.11 & 430 & 6 \\
       \feii$_{\rm ~out}$ & 9988 & 2.88$\pm$0.54 & 1834 & -279 & & 9972 & 1.02$\pm$0.53 & 1850 & -765 \\
       Pa$\delta_{\rm ~NLR}$ & 10049 & 0.17$\pm$0.04 & 137 & 0 & & 10049 & 0.10$\pm$0.03 & 137 & 0 \\
       Pa$\delta_{\rm ~BLR}$  & 10049 & 1.72$\pm$0.14 & 537 & 0 &  & 10049 & 4.08$\pm$0.14 & 598 & 0 \\
       Pa$\delta_{\rm ~out}$ & 10044 & 3.18$\pm$0.51 & 1954 & -149 & & 10026 &  1.87$\pm$0.69 & 2615 & -686 \\
       He\,{\sc ii}$_{\rm ~BLR}$ & 10120 & 0.27$\pm$1.87 & 771 & -118 & & 10120 & 0.28$\pm$0.14 &  540 & -118 \\
       \feii$_{\rm ~BLR}$ & 10172 & 0.39$\pm$0.13 & 523 & 29 & & 10172 & 0.62$\pm$0.17 & 591 & 29 \\
       
       \feii$_{\rm ~BLR}$ &   10491 & 0.40$\pm$0.16 & 415 & 0 & & 10491 & 0.50$\pm$0.14 &  427 & 0   \\
       \feii$_{\rm ~BLR}$ &  10503 & 1.63$\pm$0.16 & 414 & 57 & & 10502 & 3.04$\pm$0.14 &  432 & 29    \\
       \feii$_{\rm ~out}$    &  10489 &  2.07$\pm$0.62 &  1840 & -341 & & 10472 & 1.20$\pm$0.55 & 1856 & -826 \\
        
    O\,{\sc i}$_{\rm ~BLR}$ & 11289 & 1.97$\pm$0.15 & 440 & 53 & & 11288 &  4.60$\pm$0.47 & 546 & 27 \\
    O\,{\sc i}$_{\rm ~out}$  & 11281 & 2.27$\pm$0.58 & 1802 & -159 & & -- & -- & -- & -- \\
       Pa$\beta_{\rm ~NLR}$ &  12818 &  0.36$\pm$0.06 & 127 & 0 & & 12819 & 0.23$\pm$0.06 & 128 & 23 \\
       Pa$\beta_{\rm ~BLR}$   & 12819 &  3.52$\pm$0.20 & 543 & 23 & & 12818 & 8.20$\pm$0.21 &  602 & 0 \\
       Pa$\beta$$_{\rm ~out}$   & 12809 & 5.38$\pm$0.75 & 1945 & -210 & & 12787 & 2.70$\pm$0.84 & 2545 & -725 \\
    \hline
    \end{tabular}
    \caption{Line parameters (center, flux, FWHM, and shift from the centroid position) for the Gaussian (columns 2-5) and the Lorentzian fits (columns 6-9). Fluxes in units of 10$^{-15}$\,erg\,cm$^{-2}$\,s$^{-1}$. The subscripts ``NLR'', ``BLR'', and ``out'' refer to the component emitted by the narrow line region, the classical broad line region, and the outflow component, respectively. }
    \label{tab:line_fit}
\end{table*}

\subsection{The NIR Fe\,{\sc ii} emission} 

In addition to the permitted H\,{\sc i} lines, the NIR spectrum of \1h07\ also shows the presence of prominent \feii\ emission features, in particular, the ones at $\lambda$9997, $\lambda$10501, $\lambda$10863, and $\lambda$11127 \citep{rudy+00,rodriguez+02,marinello+20}. Because of their proximity in wavelength, they are termed as the 1~$\mu$m \feii\ lines and are emitted after the decay of the common upper term b$^4$G.  Overall, \feii\ at $\lambda$9997 and $\lambda$10501 are very conspicuous and moderately isolated lines. For that reason, they will be employed here in the analysis of the \feii\ emission line profiles. The former is $\sim$50~\AA\ apart from Pa$\delta$, the nearest strongest emission feature, while \feii\ $\lambda$10501 is indeed a blend of \feii\ at $\lambda$10493 and $\lambda$10501, rarely being resolved because of their proximity. Moreover, the expected flux ratio $\lambda10501/\lambda10493$ is $\sim$8 \citep{marinello+16}. Thus, the flux of $\lambda$10501 dominates the blend. At the redshift of \1h07, the lines at $\lambda$10863 and $\lambda$11127  are located in regions of bad atmospheric transmission, preventing us from using them reliably. 
 
\begin{figure*}[]
\includegraphics[width=9cm]{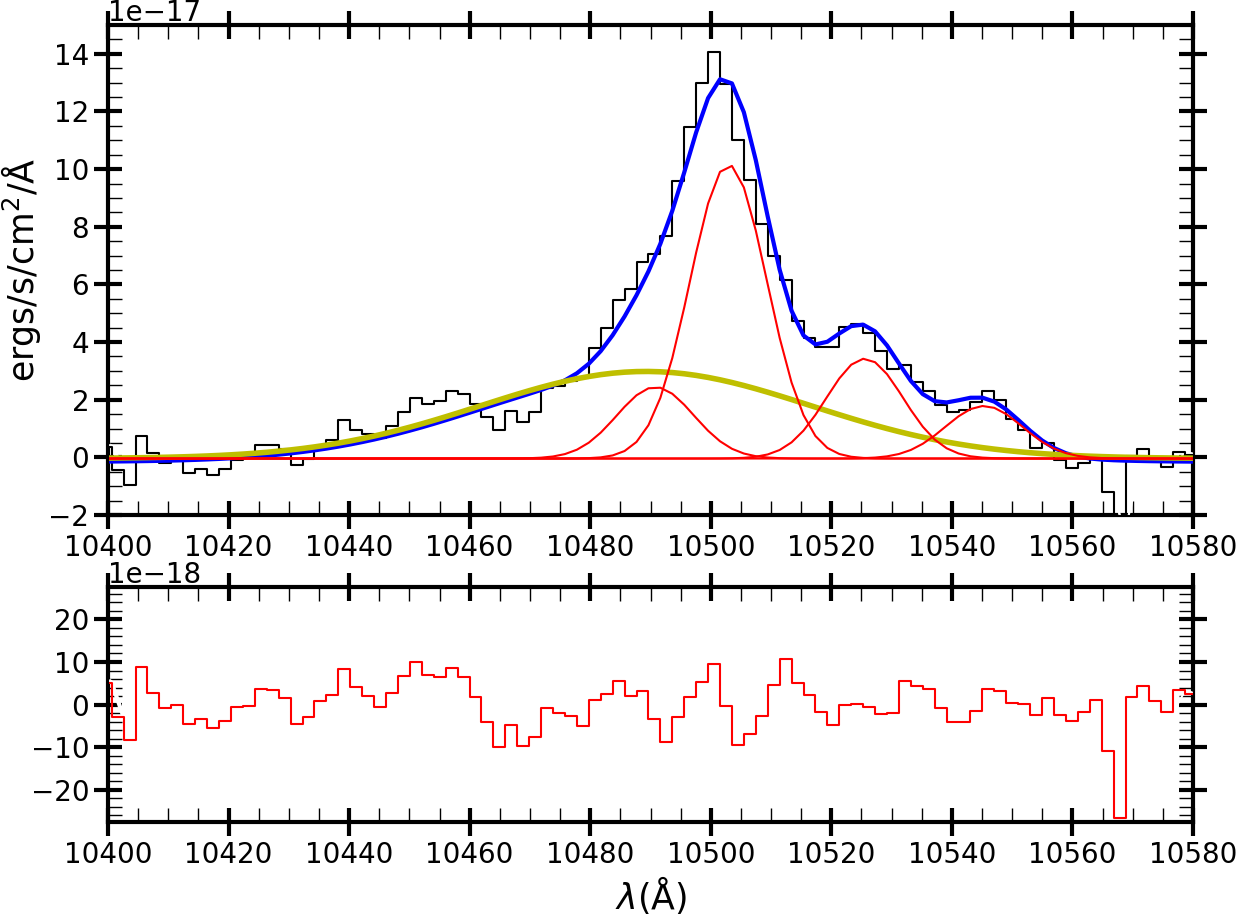}
\includegraphics[width=9cm]{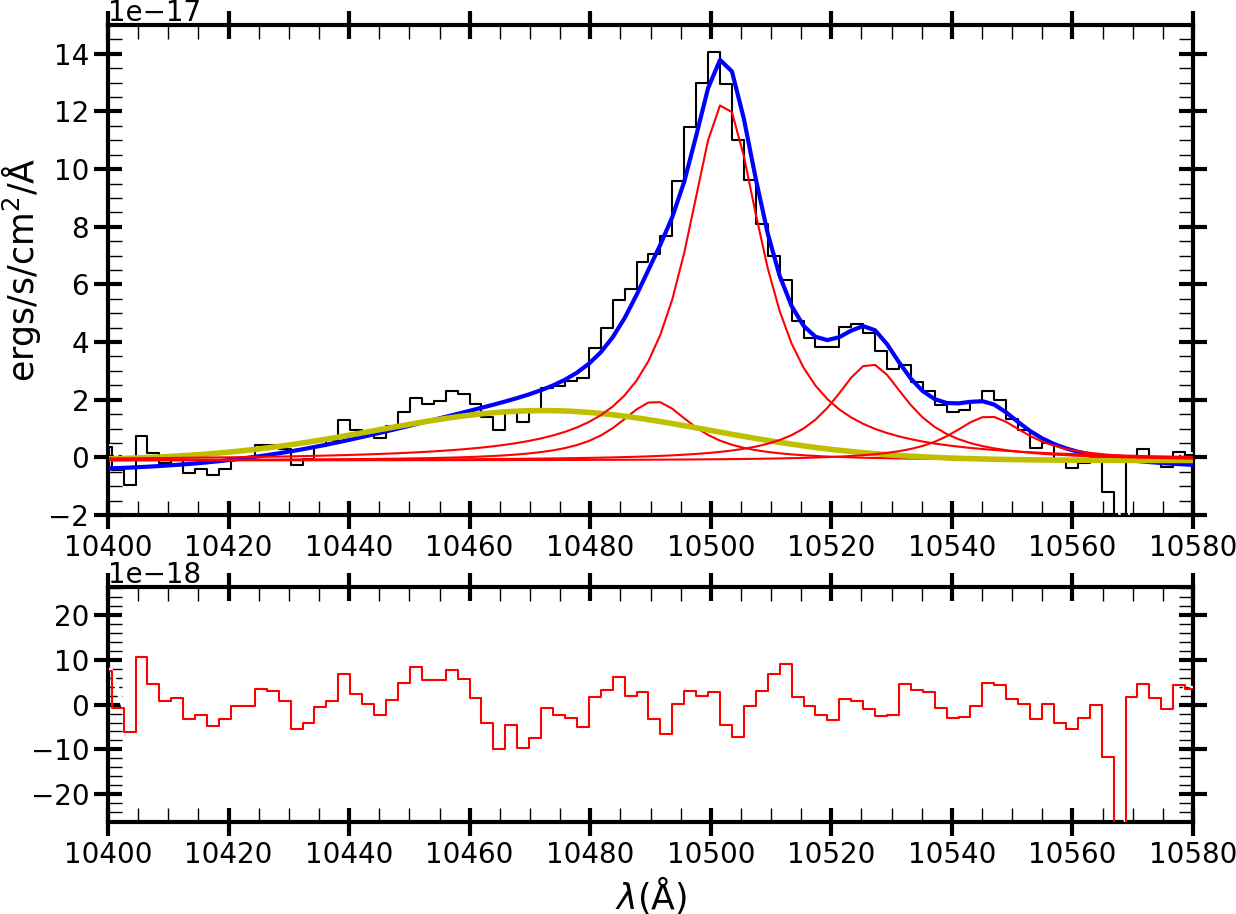}
\caption{Deblending procedure applied to the \feii~$\lambda$10501 line. The left panel shows the modeling assuming Gaussian components, while the right panel shows the results considering a Lorentzian profile for the part emitted by the classical BLR. The black histogram is the observed profile and the blue line is the fit. Individual components corresponding to the classical BLR are in red while the outflow component is in yellow. The bottom panel is the residual after subtracting from the observations the best fit. \label{fig:fe2_10500}}
\end{figure*}

In this section, we will focus on the results obtained for the Fe\,{\sc ii}$~\lambda$10501 blend. As with the Pa$\beta$ line, We first tested the hypothesis of a line profile being represented solely by the classical BLR component. The results are shown in Figure~\ref{fig:fe2_10500_no_out}. The BIC values are listed in the Table~\ref{tab:BIC_results}.
It is possible to see the presence of residuals in the blue wing in $\lambda$10501. For this reason, we included an outflow component to the most prominent line, that is, the one centered at $\lambda10501$. The left panel of Figure~\ref{fig:fe2_10500} illustrates the result after the addition of that component plus a Gaussian profile to account for the classical BLR contribution. In the right panel, a Lorentzian profile for that component is assumed. The lines at $\lambda$10493, $\lambda$10526, and $\lambda$10546 are also emitted by \feii\ but are considerably fainter features than that at $\lambda$10501. For this reason, only the classical BLR component was used to fit them. The outflow component, if present, is at the S/N level. The BIC found from these fits are in Table~\ref{tab:BIC_results}. Table~\ref{tab:line_fit} lists the best-fit parameters found for \feii{} blend centered at $\lambda$10501.

As in the \Pab\ fit, the presence of an outflow component is strongly favored. The fit with the classical BLR plus the outflow component (with the width, intensity, and centroid position left unconstrained in the fit) results in smaller BIC values, with the $\Delta$BIC being $>$ 10 when compared to that of without outflow. Moreover, the Lorentzian profile to represent the classical BLR is also favored as the BIC in the latter case is smaller than when a Gaussian profile is assumed. The $\Delta$BIC between these two cases amounts to 18.

The fits shown in Figure~\ref{fig:fe2_10500} confirm the results already gathered from Pa$\beta$. We found a classical BLR component, which within errors is characterized by a Lorentzian profile with FWHM of 432~\kms\ and coincident with the systemic velocity. This agrees with the results of \citet{leighly+04} for Mg\,{\sc ii} and \feii. Moreover, there is a broad blueshifted component that we associate with outflowing gas from the BLR. It has a FWHM of $\sim$1856 \kms\ and an outflow velocity between -341 and -826~\kms. To the best of our knowledge, this is the first time that such an outflow has been detected in \feii{} BLR emission in \1h07. It is important to notice here some additional findings. First, the \feii\ component emitted by the classical BLR displays an FWHM that is smaller than its counterpart in Pa$\beta$. This is in agreement with previous results found in the literature \citep{rodriguez+02, matsuoka+07, barth+13, marinello+16}. The latter work reports that the BLR component of \feii\ is, on average, 30\% narrower than that of Pa$\beta$. Second, the FWHM of the \feii\ lines is likely one of the smallest already reported in the literature even for a NLS1 AGN. And third, the presence of a shoulder centered at $\sim$10492, coinciding with Fe\,{\sc ii}$\lambda$10493 suggests that we detected and resolved that line. The flux ratio measured between  \feii{}~$\lambda$10501/$\lambda$10493 is 6.08$\pm$1.73 using the Lorentzian BLR model, very close to the expected ratio of $\sim$8 \citep{marinello+16}. Spectra with larger spectral resolution and S/N is necessary to confirm this result.

Finally, we have also considered the possibility of the \feii\ blend being fully dominated by the line at 10501~\AA\ while the satellite lines are spurious features. This assumption comes from the fact that the calculated intensity of these lines relative to \feii\ 10501 is at least a factor 2$\times10^{-2}$ smaller. These values result by considering only the Gaunt factors and the Einstein coefficient $A_{ij}$ of the lines. We then fit a Gaussian or a Lorentzian profile to the observed line plus the outflow component. The BIC resulting in these two cases were -5297.3 and -5298.2, respectively. That is, larger than those derived when considering the satellite lines (See~\ref{tab:BIC_results}). We highlight that the line ratios between the different \feii\ lines may significantly depart from those expected from pure recombination  because of collisional excitation, Ly-$\alpha$ fluorescence, self-fluorescence, and other processes \citep[see][]{sigut_pradhan03} that contributes to the observed \feii\ spectrum in AGN.

\subsection{The O\,{\sc i} emission line}

\begin{figure*}[]
\includegraphics[width=9cm]{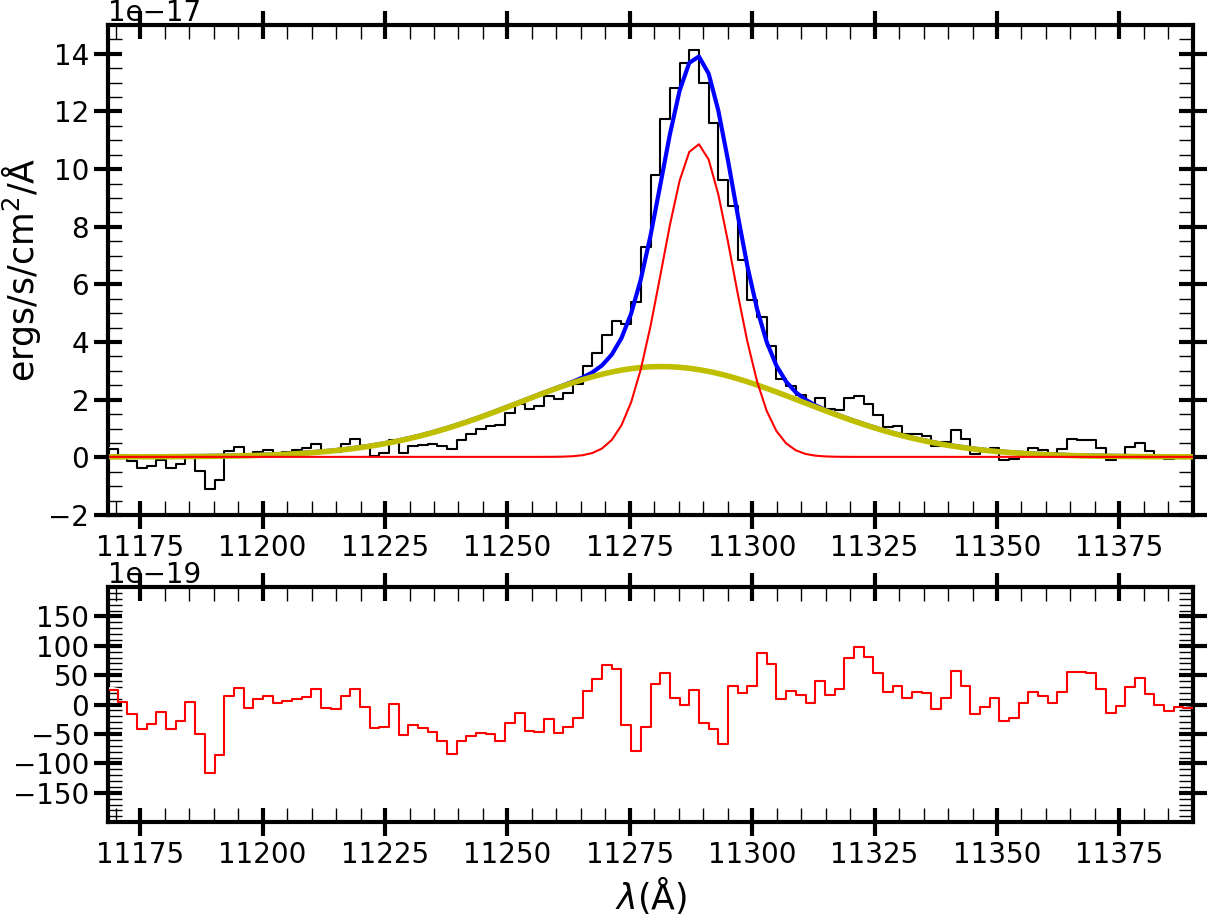}
\includegraphics[width=9cm]{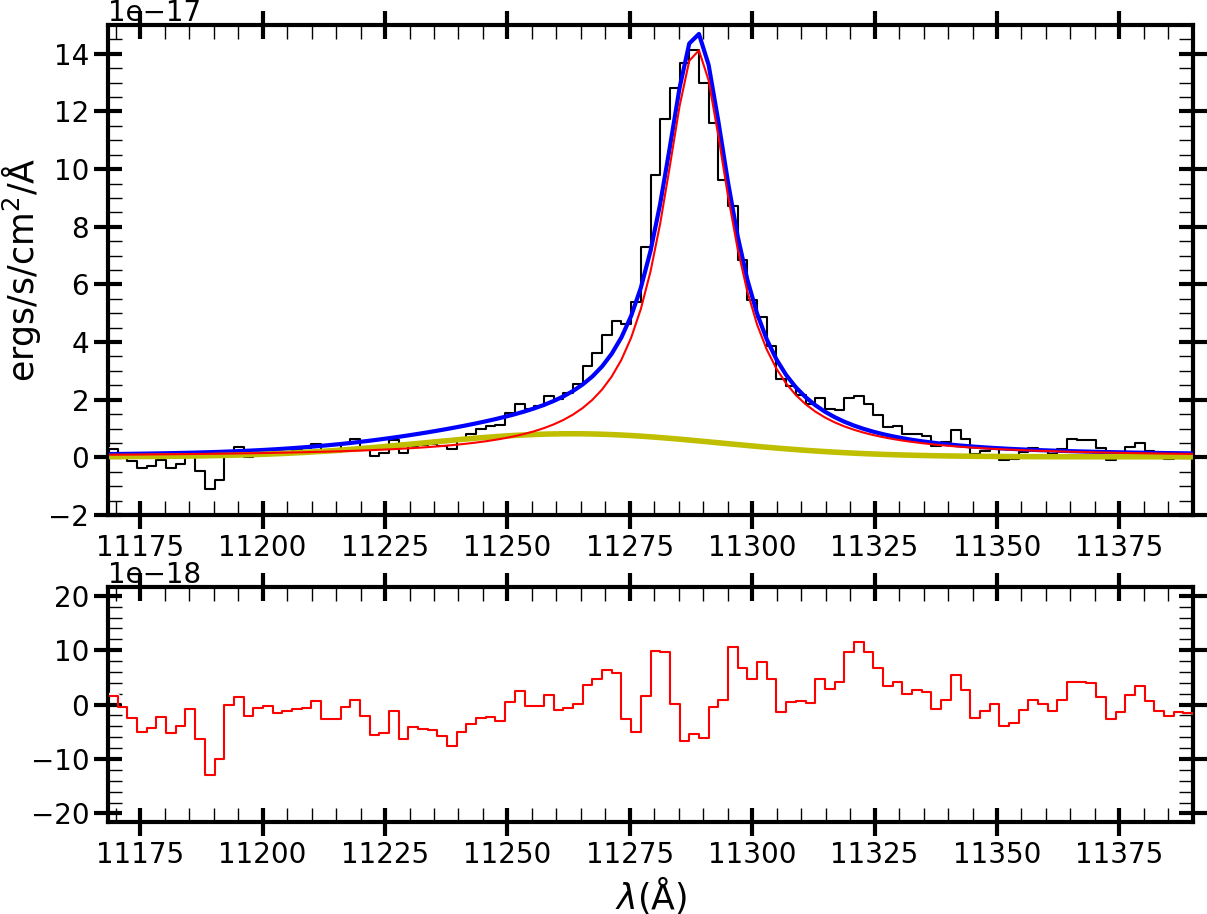}
\caption{Deblending procedure applied to the O\,{\sc i}~$\lambda$11287. The left panel shows the modeling assuming Gaussian components, while the right panel shows the results considering a Lorentzian profile for the part emitted by the classical BLR. The observed profile is the black histogram, and the blue line is the best fit.  Individual components corresponding to the classical BLR are in red while the outflow component is in yellow. The bottom panel is the residual after subtracting from the observations the best fit. \label{fig:oifit}}
\end{figure*}

We took advantage of the presence of the strong O\,{\sc i} emission at $\lambda$11287 in our spectrum and modeled that line using the two approaches already employed and described above. The fit with a single component to represent the BLR is in Figure~\ref{fig:o1_no_out} and the one including the outflow is shown in Figure~\ref{fig:oifit}. The corresponding BIC values are in Table~\ref{tab:BIC_results}. 

An inspection of Figure~\ref{fig:o1_no_out} reveals that a single Gaussian function does not reproduce satisfactorily the observed profile. Strong residuals are left in the blue and red wings of the emission line. However, when a single Lorentzian profile is employed, the residuals are considerably reduced. This is supported by the BIC, which is the smallest when that latter profile is employed.

We also tested the scenario where a blueshifted broad component is included in the fit. The results are displayed in Figure~\ref{fig:oifit} and the corresponding BIC values are listed in Table~\ref{tab:BIC_results}. It can be seen that in a classical Gaussian BLR, the outflow component produces a smaller BIC, with a difference $>$10 relative to the case of no outflow. In the case of a classical Lorentzian BLR, though, the inclusion of an outflow component does not improve the modeling. Indeed, the BIC increases considerably. Therefore, for O\,{\sc i}~$\lambda$11287 an outflow component is acceptable under the assumption that the classical BLR is represented by a Gaussian profile. Nonetheless, it is important to highlight that the smallest BIC is obtained when that latter region is modeled using a Lorentzian function, making the outflow scenario uncertain for that line. The outflow may be present but it was not detected within our S/N because of the presence of a strong telluric absorption feature to the blue side of the line.   
Table~\ref{tab:line_fit} lists the parameters found for O\,{\sc i}~$\lambda$11287 using both the Gaussian and Lorentzian approaches.

\subsection{Other line fittings}

\begin{figure}[!h]
\includegraphics[width=9cm]{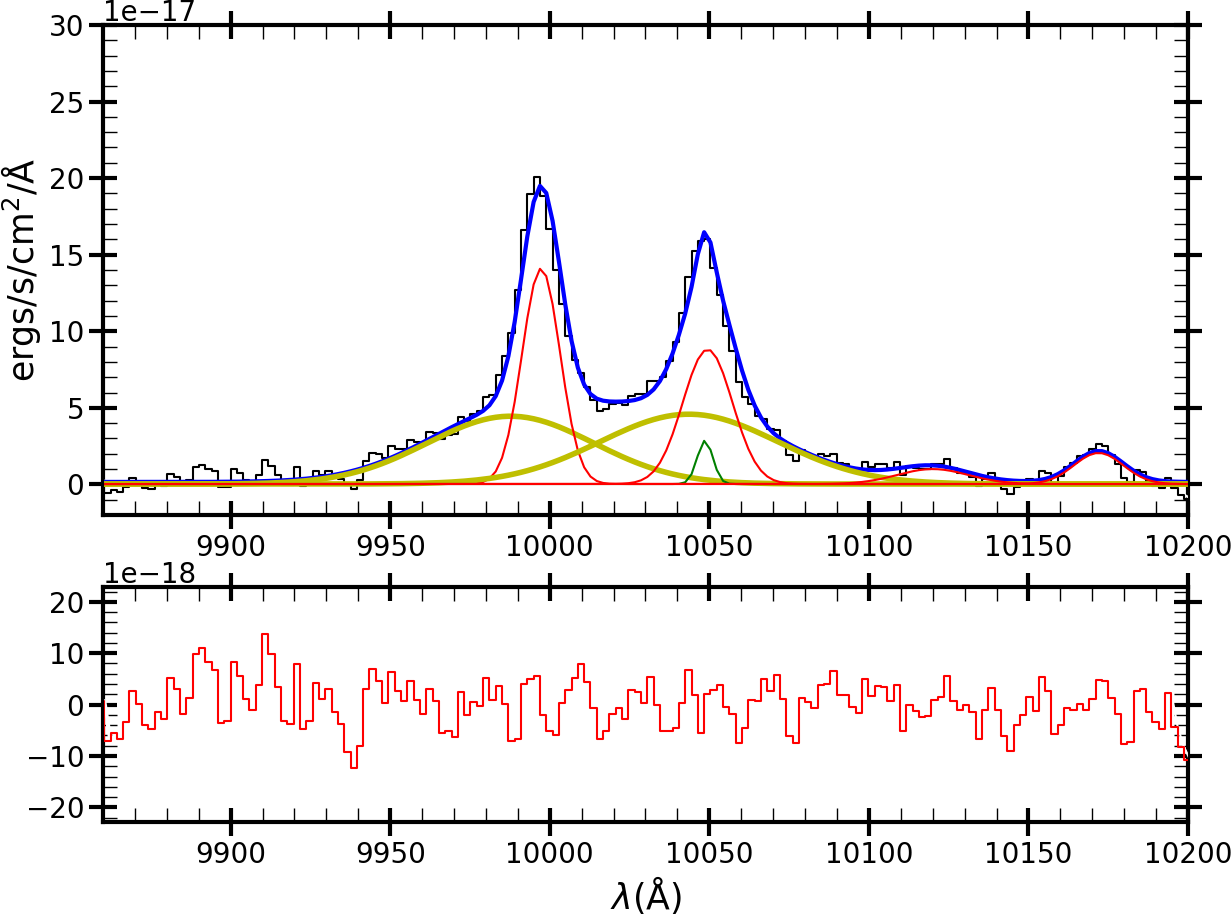}
\includegraphics[width=9cm]{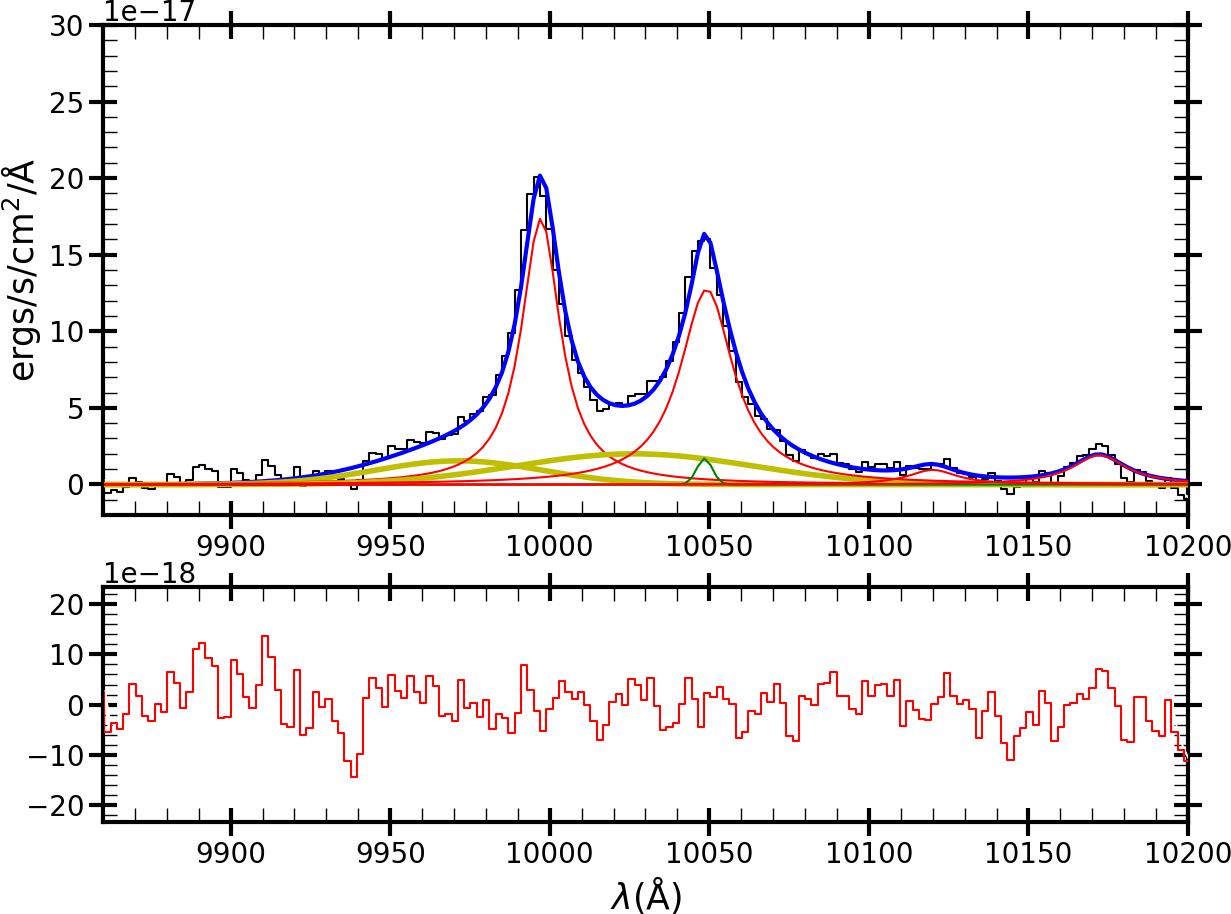}
\caption{Deblending procedure applied to the \feii~$\lambda$9997+Pa$\delta$ blend. The left panel shows the modeling assuming Gaussian components, while the right panel shows the results considering a Lorentzian profile for the part emitted by the classical BLR. The observed profile is the black histogram, and the blue line is the best fit.  Individual components corresponding to the classical BLR are in red while the outflow component is in yellow. The green line is the component attributed to the NLR. The bottom panel is the residual after subtracting from the observations the best fit. \label{fig:padelta}}
\end{figure}

We next proceed to fit the blend formed by the \feii\ lines at $\lambda$9997 and $\lambda$10172, the Pa$\delta$ line and the He\,{\sc ii} line at $\lambda$10124. Figure~\ref{fig:padelta_no_out} displays the fit when the BLR is modeled with a single component. The corresponding BIC is in Table~\ref{tab:BIC_results}. Figure~\ref{fig:padelta} illustrates the scenario where a blueshifted Gaussian component is included in addition to a Gaussian profile (left panel) and a Lorentzian profile (right panel) for the classical BLR component. In this process, we tied the Pa$\delta$ profile to have similar FWHM and centroid positions as that found for the Pa$\beta$ line. BIC values are in Table~\ref{tab:BIC_results} while Table~\ref{tab:line_fit} lists the best parameters of the different components found from the fit.

The BIC values support the need for a blueshifted broad component in both Fe\,{\sc ii} and Pa$\delta$, in agreement with the results already found using Pa$\beta$ and \feii~$\lambda$10501. Moreover, the BIC is smaller for the Lorentzian approach, indicating that the profile is the most suitable one to represent the classical BLR.

\begin{figure}[!h]
\includegraphics[width=9cm]{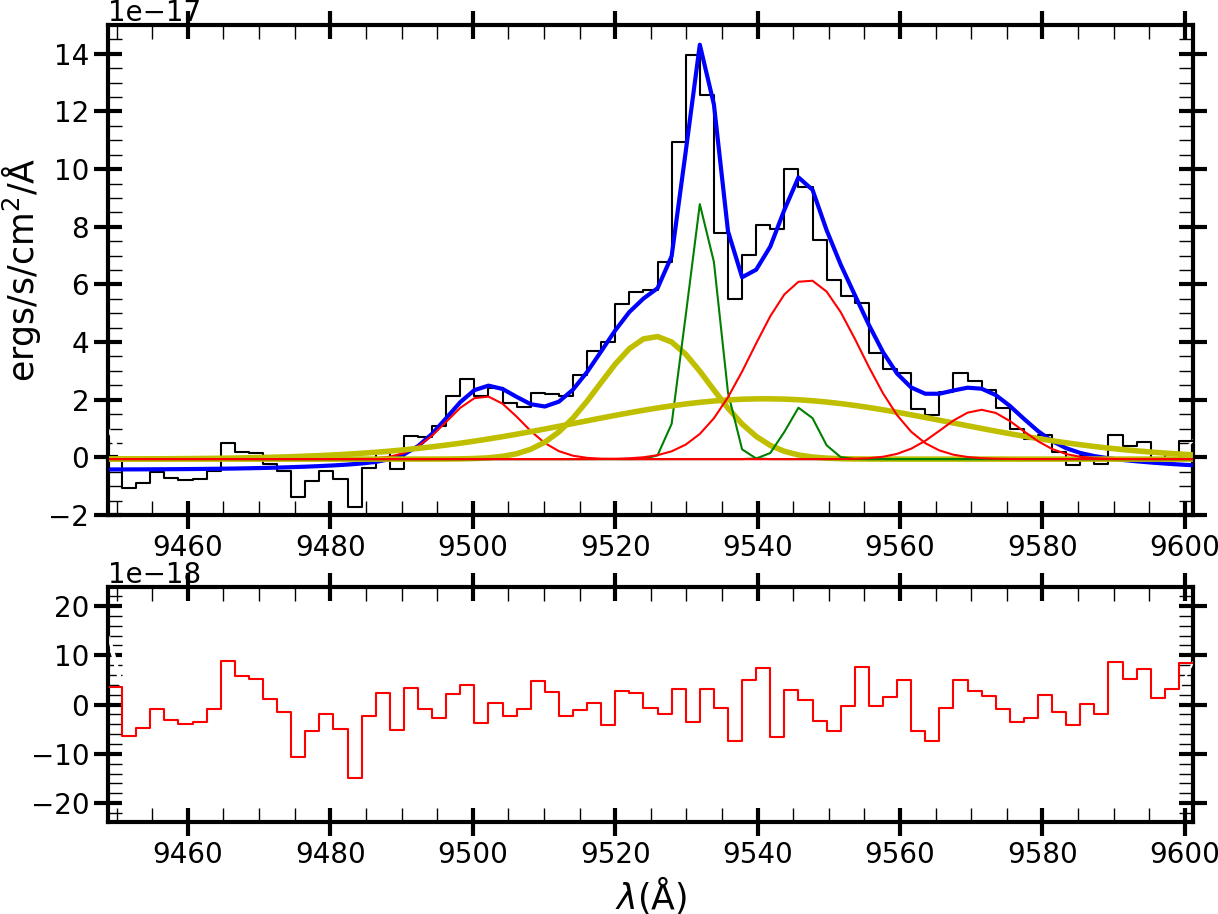}
\includegraphics[width=9cm]{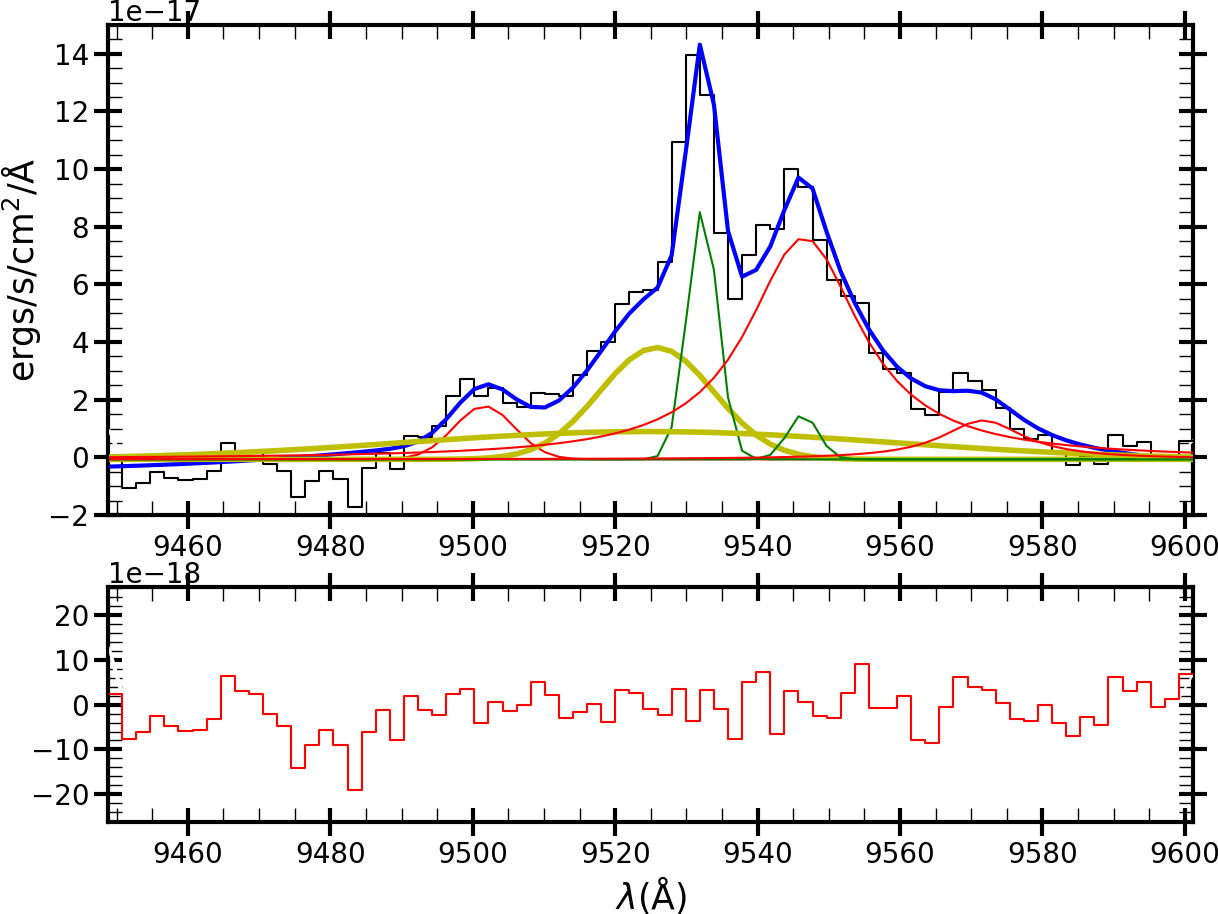}
\caption{Line fitting procedure applied to the [S\,{\sc iii}]~$\lambda$9531 + Pa~8 blend. The left panel shows the modeling assuming only Gaussian components, while the right panel shows the results considering a Lorentzian profile for the part emitted by the classical BLR. The observed profile is the black histogram, and the blue line is the modeled profile. Individual components are painted in different colors. The bottom panel is the residual after subtracting the modeled emission line profile. \label{fig:siiifit}}
\end{figure} 

Finally, we fit the blend formed by the forbidden [S\,{\sc iii}]~$\lambda$9531 line, the Pa~8 line at $\lambda$9547 and the \feii\ lines at $\lambda\lambda$9501,9571. This fit is very important because it involves the only forbidden line detected in the NIR spectrum of \1h07. Moreover, [S\,{\sc iii}]~$\lambda$9531 is considered the NIR counterpart of [O\,{\sc iii}]~$\lambda$5007 in the optical \citep{fischer+17}. 
Figure~\ref{fig:siii_no_out} pictures the result when the BLR is modeled with a single component while Figure~\ref{fig:siiifit} displays the profile fit after adding a blueshifted Gaussian component to both [S\,{\sc iii}] and Pa~8. The corresponding BIC values are in Table~\ref{tab:BIC_results}.

It can be seen that in addition to a narrow component due to the classical NLR, the [S\,{\sc iii}]~$\lambda$9531 line displays a strong blue asymmetric wing, suitably fitting with a broad component. The velocity derived from the peak centroid of the blueshifted line is -170~\kms, with a FWHM of 570~\kms\ (see Table~\ref{tab:line_fit}). Whether the blueshifted component represents the propagation of the BLR outflow into the NLR cannot be confirmed or discarded from the data. However, it is rather tempting to assume that it is, considering that the BLR lines thought to be produced in the outskirts of that region do display an outflow component.  The BIC listed in Table~\ref{tab:BIC_results} shows that the smallest values are obtained when a blueshifted component is added to the blend, confirming the results discussed above. Moreover, the BIC$_{\min}$ criterion favors the Lorenztian profile for the classical BLR.

%%%%%%%%%%%%%%%%%%%%%%%%%%%%%%%%%%%%%%%%%%
\section{Discussion}\label{discussion}

Ultrafast outflows (UFOs) are the most extreme subset of AGN winds with velocities greater than 10000~\kms. They are believed to originate from the inner accretion disk within a few hundred gravitational radii from the black hole \citep{tombesi+10,nardini+15}. The existence of an UFO has been confirmed in \1h07\ \citep{dauser+12,hagino+16,kosec+18} at a velocity of $\sim$0.13c with an ionisation parameter log($\xi$/erg\,cm\,s$^{-1}$) $\sim$4.3. \citet{blustin+09} detected Doppler-shifted emission lines, and \citet{kosec+18} found that the velocities of the blueshifted emission increase in higher ionization species. This implies that the wind in \1h07\ is likely stratified and is perhaps slowing down and cooling at larger distances from the SMBH.

The evidence gathered in this work using NIR spectroscopy and BIC analysis reveals the presence of a blueshifted emission in the low-ionization lines of H\,{\sc i} (Pa$\beta$, Pa$\delta$, and Pa~8), Fe\,{\sc ii} $\lambda\lambda$9997,10501, and, and to a lower extend, in O\,{\sc i}~$\lambda$11287. Even the forbidden line of [S\,{\sc iii}]~$\lambda$9531 displays clear evidence of such outflow.  In all cases, the outflow component is modeled using a Gaussian function with a FWHM between 1800 and 2600~\kms.  The outflow velocity varies from -160~\kms\ to -820~\kms\ in the lines studied. The lowest velocity is derived when a Gaussian function is employed to model the classical BLR contribution, while the largest one is obtained when a Lorentzian profile is used instead.

We suggest that the outflow that is detected in the BLR high-ionization lines reported in \citet{leighly+04} is also present in the low-ionization lines of H\,{\sc i} and \feii{}. Previous observations of low-ionization lines such as Mg\,{\sc ii}, in the UV, and the Balmer lines, in the optical, made by the latter authors revealed that they appear to be rest-frame and of disc origin. Here, we confirm the detection of such component using BLR NIR lines, at rest relative to the systemic velocity and with a turbulent velocity (FWHM) between 410 and 600~\kms. Moreover, the outflow component is unambiguous, regardless of the type of profile employed in the line fitting to represent the classical BLR although our study favours the Lorentzian representation. The outflow velocity can be as high as $\sim$826~\kms.  

\citet{kosec+18} found a trend of increasing velocities with the increase of the ionization parameter of the ions. They also report possible similar kinetic powers of UV, soft X-ray emitters, and UFO absorbers, suggesting that we are witnessing the evolution of a stratified, kinetic energy-conserving wind. Our results point out that the wind in \1h07\ extends well into the outer boundary of the BLR, where the \feii\ is formed \citep{barth+13, marinello+16}.  
Indeed, adding the \feii\ measurements from this work to Figure~3 of \citet{kosec+18}, which relates the strength of the outflows with the ionisation parameter, gives further support to our claims. They found a relationship of the form $V = b \times (\xi$ /erg\,cm\,s$^{-1})^a$, where $V$ is the outflow velocity, $\xi$ is the ionisation parameter,
and $a$ and $b$ are constants, $a$ = 0.36$\pm$0.04 and $b$ = 1800 $\pm$ 300~\kms. 

To derive the expected outflow velocity for Fe {\sc ii} we employed log~$U$ of $\approx$ -3.25. This value is calculated using {\sc cloudy} \citep{ferland+17} and the observed SED for I\,Zw\,1, super-solar metal content (5-10 Z$_{\odot}$), and a microturbulence velocity of 20~km\,s$^{-1}$ within the BLR cloud. The model further assumes a cloud column density of 10$^{24}$ cm$^{-2}$ \citep[see][for more details]{2021A&A...650A.154P}. From the above equation and employing the aforementioned value of $U$, an outflow velocity of -125~\kms\ is predicted. We note that \citet{kosec+18} choose arbitrary 500~\kms\ error bars on UV ion velocities due to a lack of uncertainties in \citet{leighly+04}. Thus, assuming this uncertainty, the outflow velocity measured here in \feii{} and H\,{\sc i} is less than a factor 2 than the one predicted. \ion{O}{1} may still be taking part of the outflow, as the Gaussian approach for the classical BLR suggests. Using the same {\sc cloudy} model as for \feii{}, a log~$U$ $\approx$ -3 is predicted. The corresponding outflow velocity is -222~\kms\ while we measured a value of -159$\pm$30~\kms. We remark that the blue wing of this line may be affected by telluric absorption, hindering its detection using the Lorentzian BLR scenario. The detection of a blueshifted component in the forbidden line of [\ion{S}{3}] indeed suggests that the outflow is coupled to the gas located in the NLR.  

Another important result is that the disc component of the low-ionization lines displays a very low turbulent velocity, of $\sim$546~\kms\ in the \feii\ and the O\,{\sc i} lines. These are likely some of the narrowest BLR lines already reported in the literature.  

A full analysis of the NIR spectrum of \1h07\ is out of the scope of this work. However, upon inspection of our spectrum, it is evident that IFU observations of this object, both in the optical and in the NIR, are necessary to fully assess the size and energetic properties of the outflow. 

%%%%%%%%%%%%%%%%%%%%%%%%%%%%%%%%%%%%%%%%%%
\section{Conclusions}\label{conclusion}

We have analyzed the NIR spectrum of the xA NLS1 galaxy \1h07, widely known for displaying a UFO from X-ray emission lines. The ultra-fast outflow is likely propagated to the inner portion of the BLR because UV high- and mid-ionization lines produced in that region are strongly blue asymmetric. In this work we provide convincing evidence that the outflow is observed farther out, reaching the outer portions of the BLR. Our results also suggest that at least part of the outflow is detected in the NLR because of the blue asymmetric line associated with [S\,{\sc iii}]~$\lambda$9531. 

The permitted emission lines studied in this work (i.e., H\,{\sc i}, \feii, and O\,{\sc i}) are thought to be formed in the mid and outer portions of the BLR. The analysis of the line profiles shows that overall, the classical BLR component is best represented using a Lorentzian profile plus and outflow component of up to $\sim$-730~\kms\ for H\,{\sc i} and \feii. From our modelling, O\,{\sc i} seems not to take part in the outflow.  This result is supported by the use of the Bayesian Information Criteria (BIC) to select the model that best represents each line. Previous observations have failed to detect the outflow component in low-ionization lines in this AGN. Our results are also consistent, within the uncertainties, to models predicting outflow velocities down to -300~\kms, for lines formed in gas with $U<$\,${-3}$, as is the case here.

Finally, the analysis made on the NIR continuum of \1h07\ shows that it is well represented by the low-energy tail of the optical power-law, with a similar spectral index of $\alpha$=1.95. Moreover, we found strong evidence of the presence of hot dust, with a temperature of $\sim$1300~K. This dust very likely is located in the inner face of the torus. 

%%%%%%%%%%%%%%%%%%%%%%%%%%%%%%%%%%%%%%%%%

%% Also note that the akcnowlodgment environment does not support long amounts of text. If you have a lot of people and institutions to acknowledge, do not use this command. Instead, create a new \section{Acknowledgments}.
\begin{acknowledgments}
The authors thank to the anonymous referee for comments/suggestions to this manuscript. We thank the Brazilian Agencies: Conselho Nacional de Desenvolvimento Cient\'{\i}fico e Tecnol\'ogico (CNPq) and Agency of Coordenação de Aperfeiçoamento de Pessoal de Nível Superior (CAPES).
\end{acknowledgments}


\vspace{5mm}
\facilities{SOAR Telescope}


\software{Matplotlib \citep{hunter2007}, 
        Numpy \citep{van2011},
        Scipy \citep{virtanen2020},
          Cloudy \citep{ferland+17}
          }



\appendix
\section{Additional Fitting plots}

\begin{figure}
\includegraphics[width=9cm]{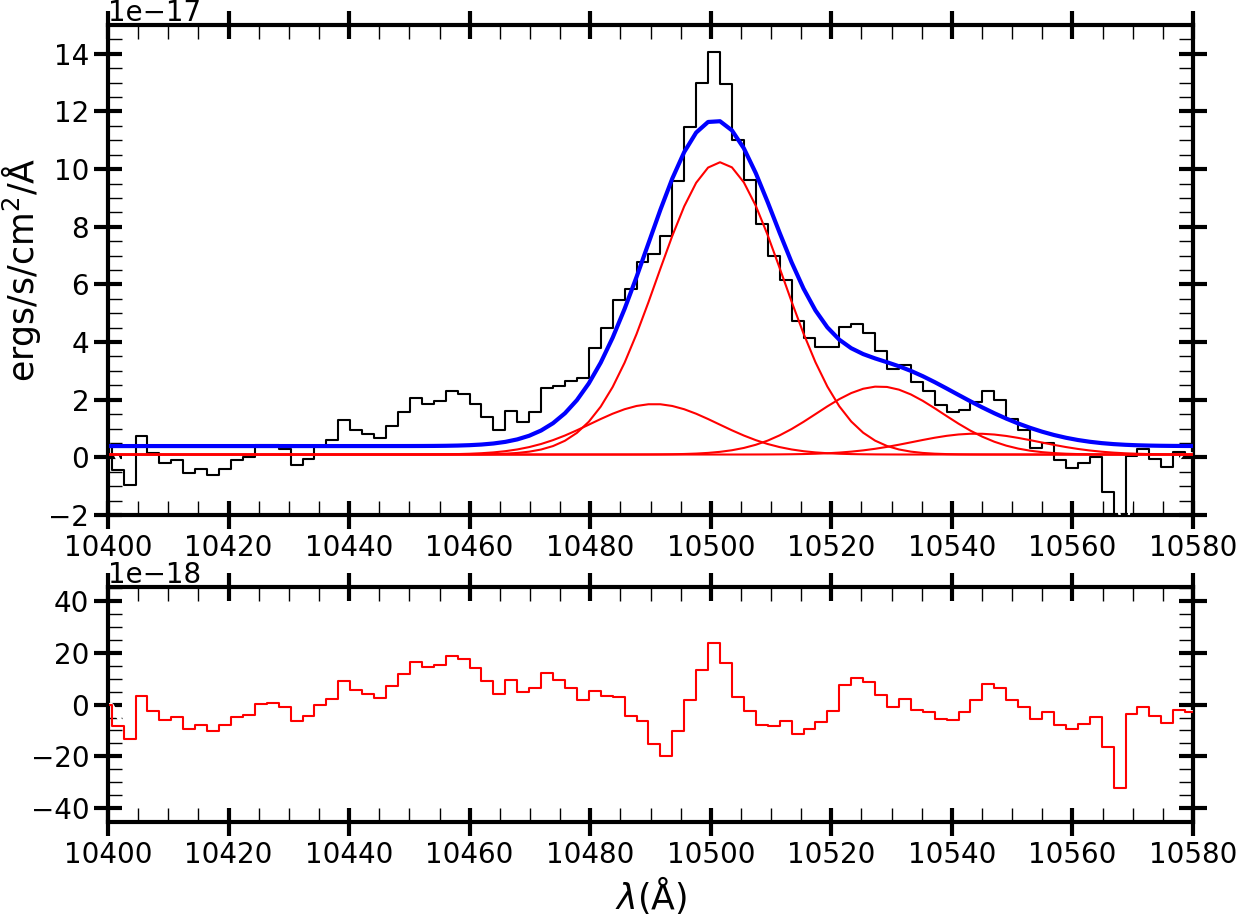}
\includegraphics[width=9cm]{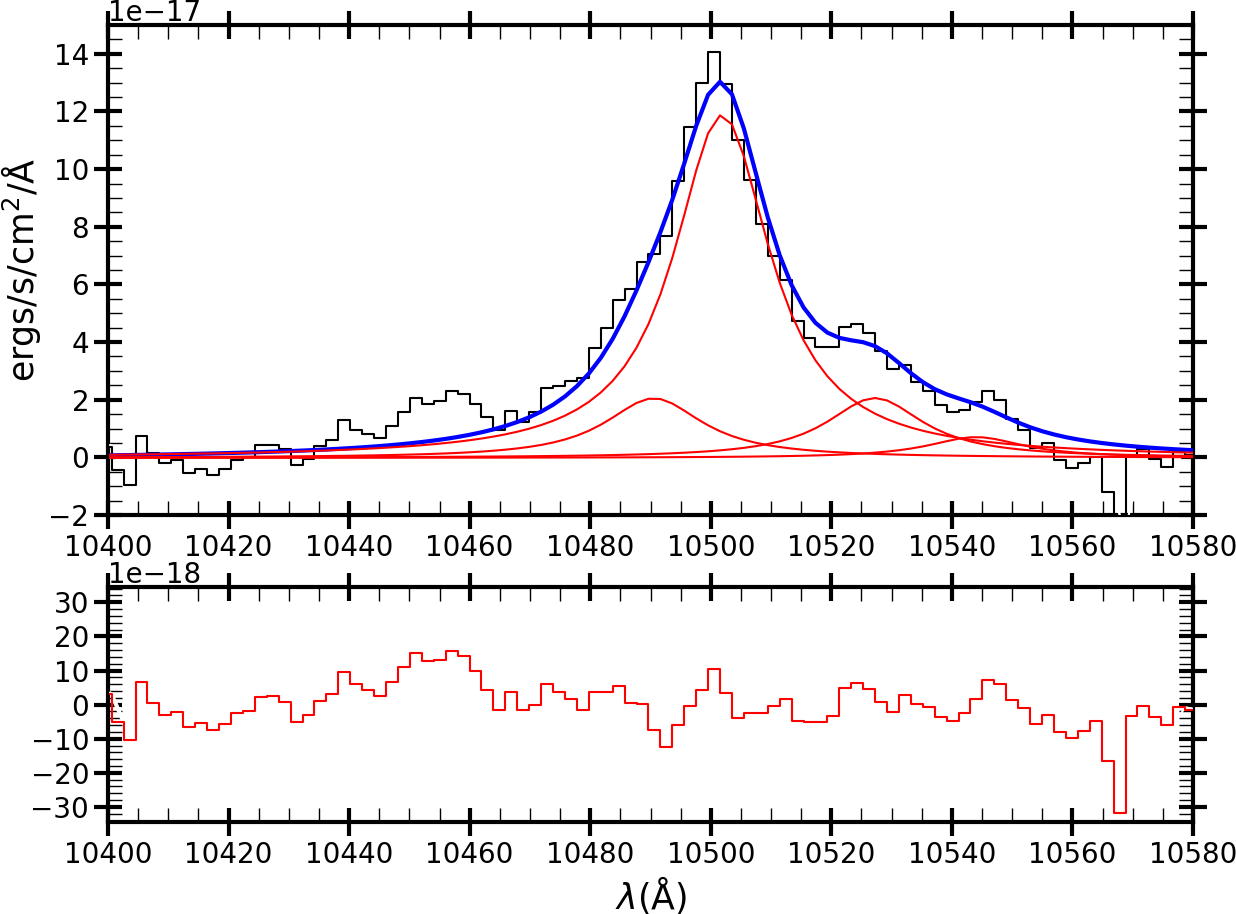}
\caption{Line fit carried out to the Fe\,{\sc ii}~$\lambda$10501 lines. The left panel shows the results after considering a BLR composed of a single Gaussian while the right panel the fit when a Lorentzian component is assumed. The BIC found from these fits are in Table~\ref{tab:BIC_results}.  The observed profile is the black histogram, and the red curve is the BLR contribution. The blue line is the modeled profile. The bottom panel is the residual after subtracting the modeled emission line profile from the observations.   \label{fig:fe2_10500_no_out}}
\end{figure}

\begin{figure}
\includegraphics[width=9cm]{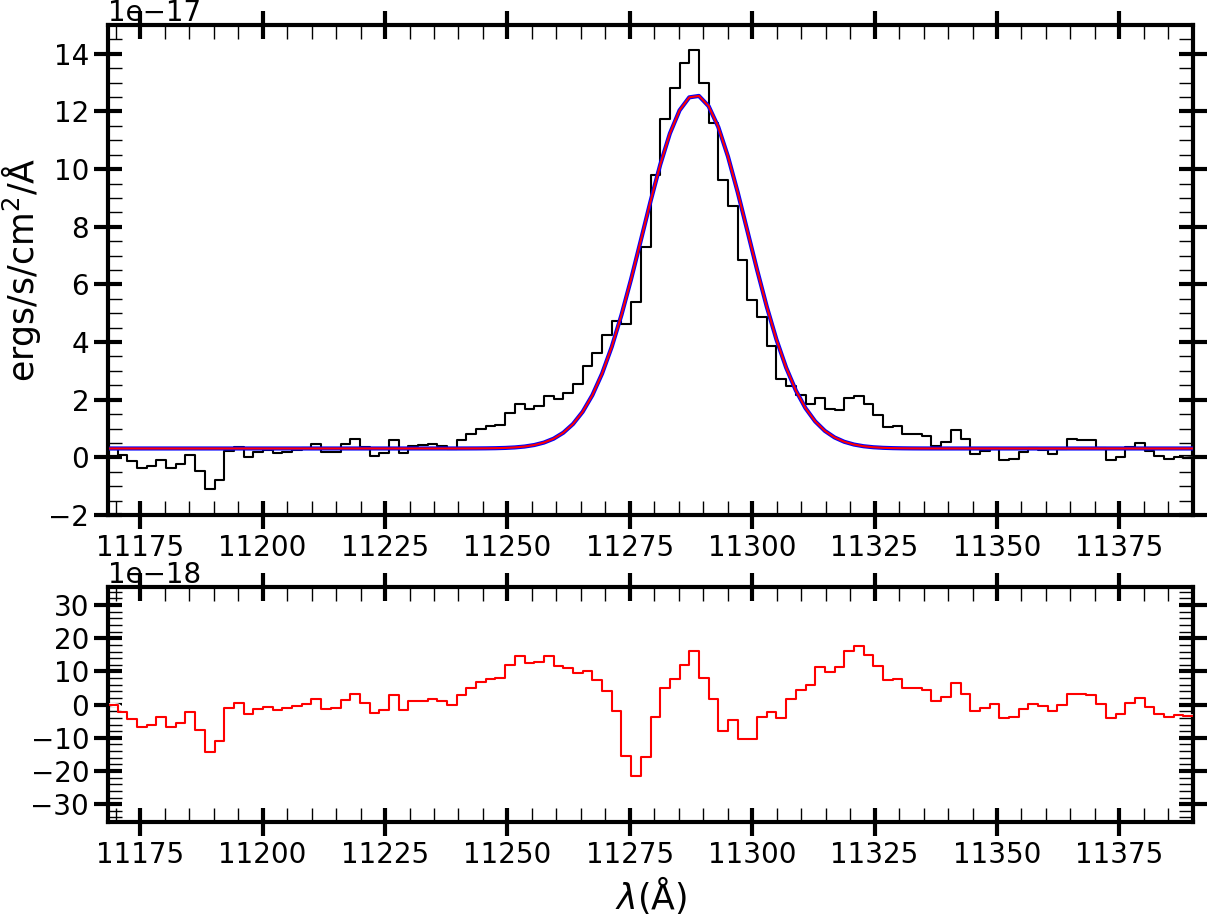}
\includegraphics[width=9cm]{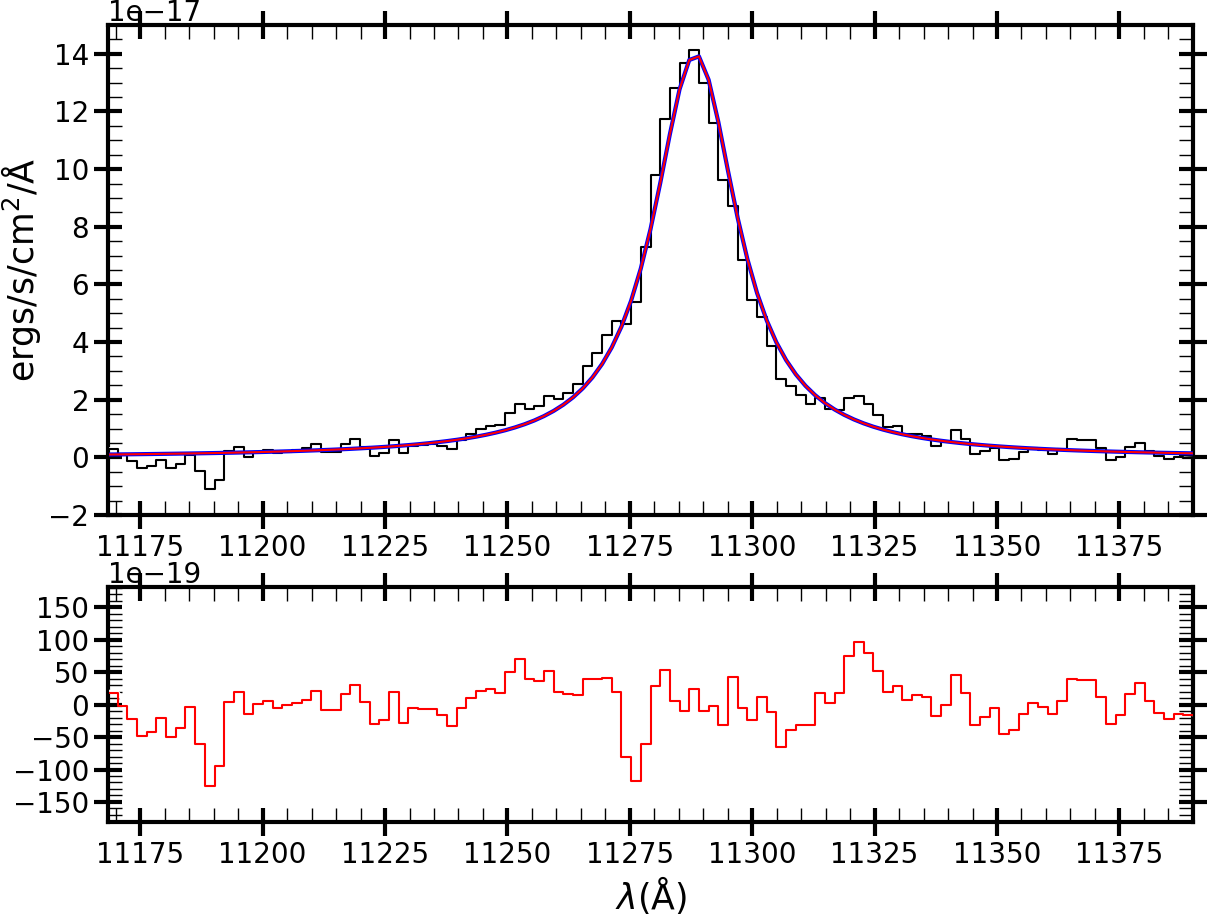}
\caption{Line fit carried out to O\,{\sc i}~$\lambda$11287. The left panel shows the results after considering a BLR composed of a single Gaussian while the right panel the fit when a Lorentzian component is assumed. The BIC found from these fits are in Table~\ref{tab:BIC_results}.  The observed profile is the black histogram, and the red curve is the BLR contribution. The blue line is the modeled profile. The bottom panel is the residual after subtracting the modeled emission line profile from the observations.   \label{fig:o1_no_out}}
\end{figure}  

\begin{figure}
\includegraphics[width=9cm]{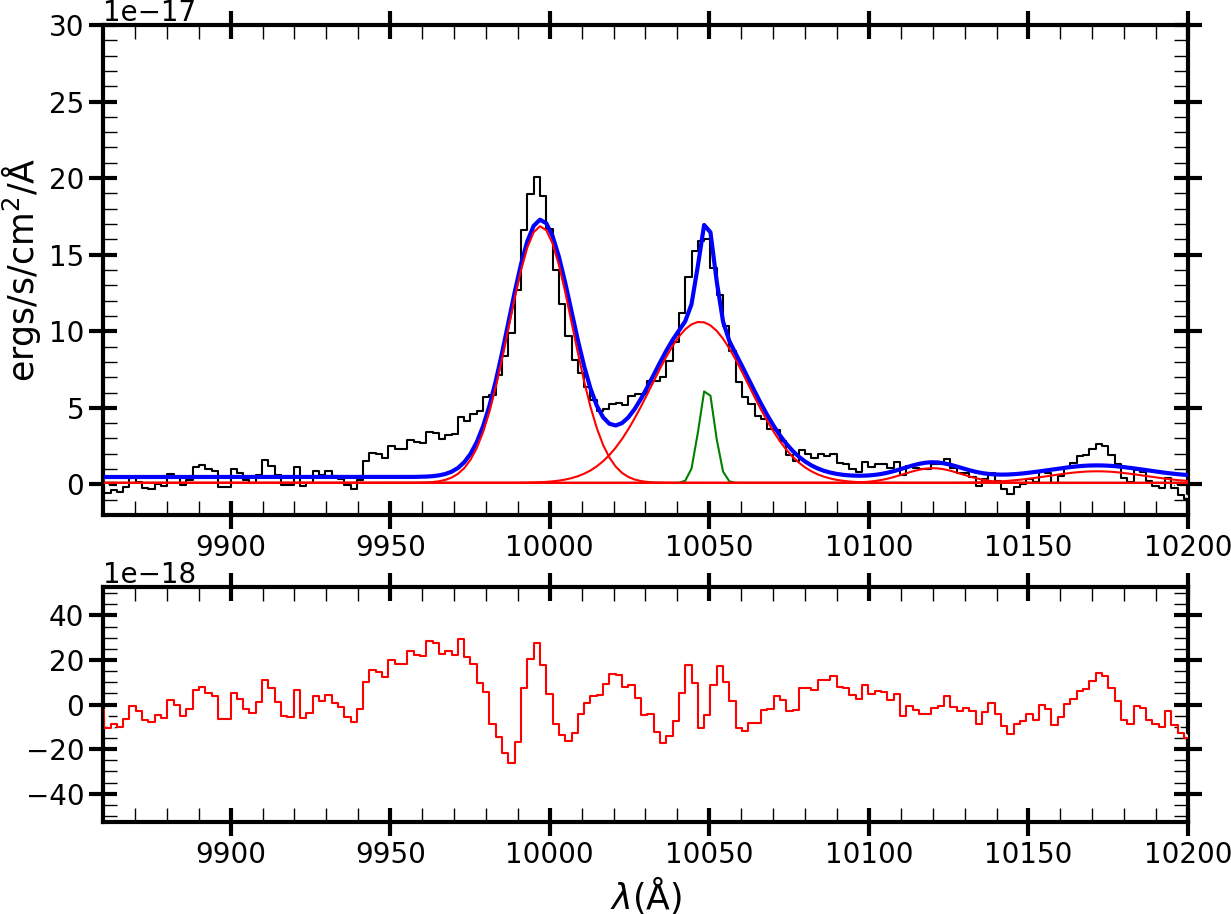}
\includegraphics[width=9cm]{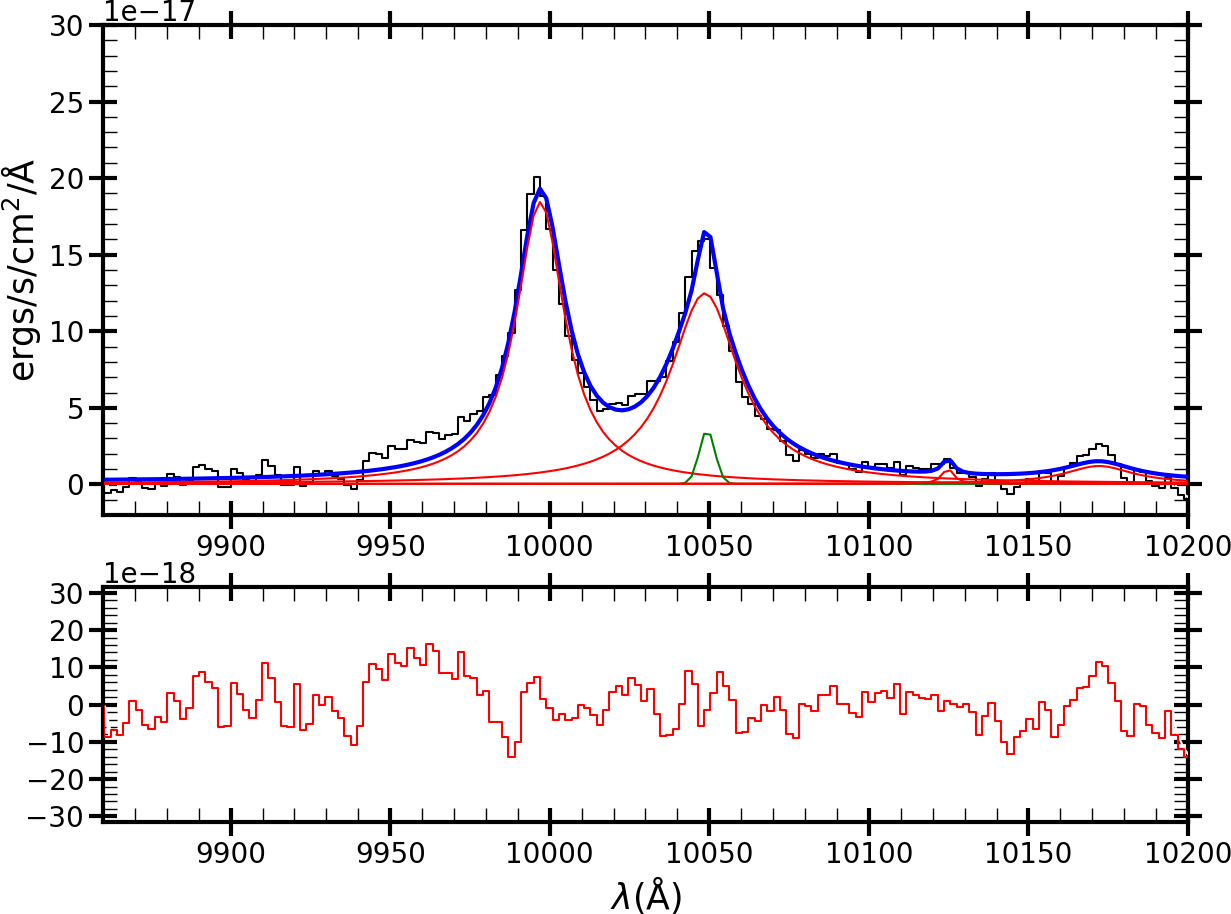}
\caption{Deblending procedure applied to the \feii~$\lambda$9997+Pa$\delta$ blend without an outflow component. The left and right panels show the results assuming that the classical BLR is represented by a Gaussian and a Lorentzian profile, respectively. The observed profile is the black histogram, and the blue line is the best fit.  Individual components corresponding to the classical BLR are in red. The green line is the component attributed to the NLR. The bottom panel is the residual after subtracting from the observations the best fit. \label{fig:padelta_no_out}}
\end{figure}

\begin{figure}
\includegraphics[width=9cm]{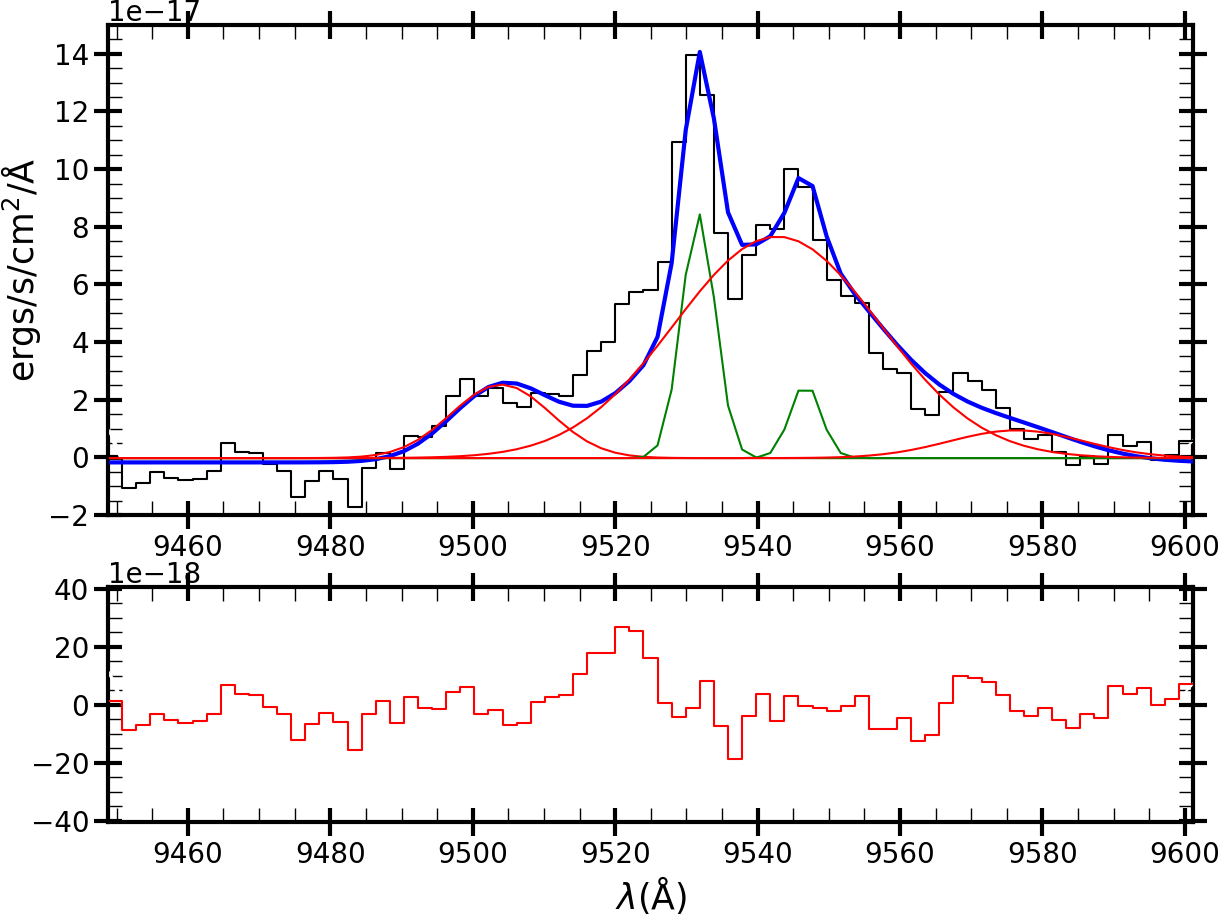}
\includegraphics[width=9cm]{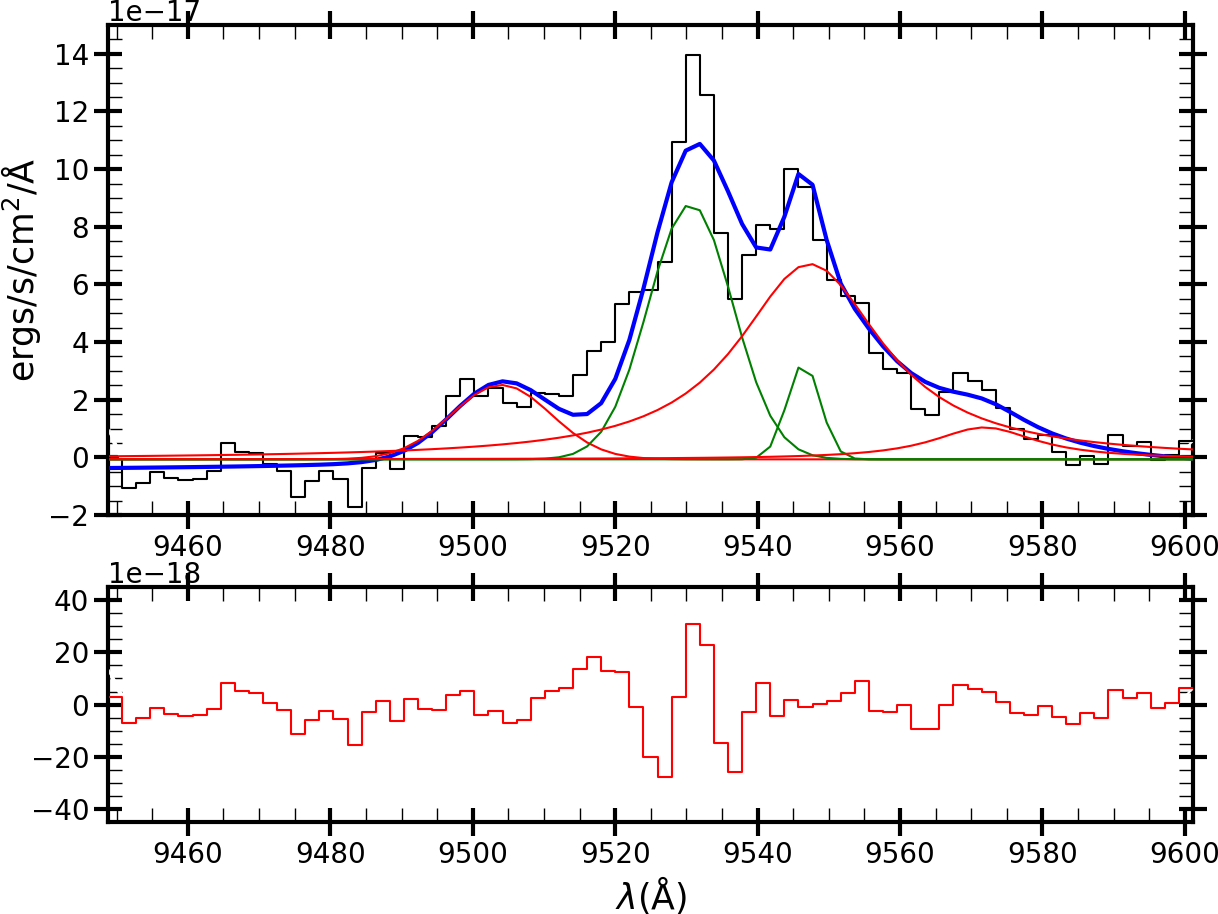}
\caption{Similar to Figure~\ref{fig:padelta_no_out} for the [S\,{\sc iii}]~$\lambda$9531+Pa~8 blend. \label{fig:siii_no_out}}
\end{figure}

%\section{Appendix information}


\bibliography{references}{}
\bibliographystyle{aasjournal}



\end{document}